\documentclass[11pt]{elsarticle}      
\usepackage[english]{babel}
\usepackage{amsmath}
\usepackage[toc,page]{appendix}
\usepackage{amssymb}        
\usepackage{amsthm}         
\usepackage{natbib}         
\usepackage{graphicx}       
\usepackage{color}
\usepackage[export]{adjustbox}
\usepackage[ruled]{algorithm2e}
\usepackage{tabularx, booktabs, multirow}
\usepackage{listings}
\usepackage{color}
\DeclareFixedFont{\ttb}{T1}{txtt}{bx}{n}{12} 
\DeclareFixedFont{\ttm}{T1}{txtt}{m}{n}{12}  

\usepackage{color}
\definecolor{deepblue}{rgb}{0,0,0.5}
\definecolor{deepred}{rgb}{0.6,0,0}
\definecolor{deepgreen}{rgb}{0,0.5,0}
\definecolor{red}{rgb}{1,0,0}
\usepackage{listings}

\newcommand\pythonstyle{\lstset{
language=Python,
basicstyle=\ttm\tiny,
otherkeywords={self},             
keywordstyle=\ttb\tiny\color{deepblue},
emph={MyClass,__init__},          
emphstyle=\ttb\tiny\color{deepred},    
stringstyle=\color{deepgreen},
frame=tb,                         
showstringspaces=false ,           %
commentstyle=\color{red},
breaklines=true
}}

\usepackage[abs]{overpic}
\usepackage{pict2e}

\lstnewenvironment{python}[1][]
{
\pythonstyle
\lstset{#1}
}
{}

\newcommand\pythoninline[1]{{\pythonstyle\lstinline!#1!}}
\newcommand{\rev}{}

\begin{document}

\begin{frontmatter}
\title{An FFT-based approach for Bloch wave analysis: application to polycrystals}
\author{Javier Segurado$^{1, 2}$\corref{cor1}}
\author{Ricardo A. Lebensohn$^{3}$}
\address{$^1$ Department of Materials Science, Technical University of Madrid, 28040 - Madrid, Spain.  \\  
               $^2$ IMDEA Materials Institute, 28906, Getafe, Madrid, Spain. \\ 
               $^3$ Los Alamos National Laboratory, Los Alamos, NM 87545, USA.}
\cortext[cor1]{Corresponding author}
\begin{abstract}

A method based on the Fast Fourier Transform is proposed to obtain the dispersion relation of acoustic waves in heterogeneous periodic media with arbitrary microstructures. The microstructure is explicitly considered using a voxelized Representative Volume Element (RVE). The dispersion diagram is obtained solving an eigenvalue problem for Bloch waves in Fourier space. To this aim, two linear operators representing stiffness and mass are defined through the use of differential operators in Fourier space. The smallest eigenvalues are obtained using the implicitly restarted Lanczos and the subspace iteration methods, and the required inverse of the stiffness operator is done using the conjugate gradient with a preconditioner. The method is used to study the propagation of acoustic waves in elastic polycrystals, showing the strong effect of crystal anistropy and polycrystaline texture on the propagation. It is shown that the method combines the simplicity of classical Fourier series analysis with the versatility of Finite Elements to account for complex geometries proving an efficient and general approach which allows the use of large RVEs in 3D.

\end{abstract}
\end{frontmatter}

\section{Introduction}

Many heterogeneous materials present certain level of periodicity associated with their microstructure, from strictly periodic, to quasi-periodicity associated with a characteristic length linked with changes in local mechanical properties. The most obvious case are additively manufactured architectural metamaterials \cite{YU2018114}, in which the microstructure is formed by the repetition of a small unit cell which \rev{conferees} special mechanical macroscopic properties, such as auxetic response or high energy absorption capacity. Other materials, like long fibre composites, foams, or some biomaterials also exhibit periodicity that might not be geometrically exact, but which allows approximating their microstructure by periodic repetition of a unit cell. On the other hand, \emph{random} materials \cite{T01} such as particle reinforced composites or polycrystals cannot be described as the repetition of a unit cell, but also present periodic change in mechanical properties at a given length scale. For example, in polycrystals with uniform or very narrow grain size distribution, a single crystal grain can be considered approximately as a unit which forms the polycrystal by some regular arrangement but, since each grain might have different orientation, this simple unit cell cannot serve to fully represent the material.

In all cases, this microstructural periodicity plays a fundamental role in the propagation of elastic waves. Material systems with strong periodicity are sometimes called phononic crystals, and the interaction of elastic waves with the microstructure might lead to the appearance of band gaps, defined as frequency ranges for which propagation of mechanical waves is hindered. This property is actively used to design such materials as wave guides or for cloaking applications \cite{cummer_nature2016}. In the cases of quasi-periodic materials or random hetergeneous materials with a characteristic length, elastic waves also present dispersive effects due to the interaction with the microstructure. These effects are not reflected in the presence of band-gaps, but introduce a dependency of the wave group speed with the frequency that induces a reduction of the wave speed for wavelengths near the microscopic characteristic length \cite{pamel2017}.

The study of wave propagation in periodic media was pioneered by Bloch \citep{Bloch1929} to describe the quantum wave functions of electrons in a crystal lattice. The Bloch-Floquet formalism establishes that for any field $u$ traveling as an harmonic wave in a periodic infinite medium with periodicity $L$, the value of the field at a point $x\in(-\infty,\infty)$ should follow
\begin{equation*}\label{eq:Bloch_wave}
u(x,t)=U(x)\mathrm{e}^{\mathrm{i}kx-\mathrm{i} \omega t}
\end{equation*}
where $U(x)$ is a periodic function with periodicity $L$, $U(x+n L)=U(x)$ for $n\in\mathbb{Z}$, and $k$ and $\omega$ correspond to the wave number and frequency respectively. 

In the case of continuum mechanics, $u$ corresponds to the displacement field and introducing this wave form in the acoustic wave equation leads to an eigenvalue differential problem in the periodic domain, whose resulting eigenvalues for a given incident $k$ correspond to the discrete number of possible $\omega$'s. The resolution of this eigenvalue problem in a range of $k$'s allows obtaining the dispersion relation $\omega(k)$ which defines the actual sound speed for each $k$, the group velocity, defined as 
\begin{equation*}\label{eq:wave_group}
c=\frac{\mathrm{d}\omega}{\mathrm{d}k}.
\end{equation*}
Due to its relevance in different fields, several numerical methods have been proposed in the last 50 years to solve this eigenvalue problem to obtain the dispersion diagram. The most common ones are the Fourier series (FS) or plane wave analysis, the Multiple Scatter Theory (MST) and the Finite Elements method (FEM), all of which will be briefly described below. Other common numerical approaches to solve partial differential equations have also been considered, like finite differences, fast multipoles, or wavelets \cite{Hussein2009}, while other common approaches as FFT-based homogenization remain almost unexplored in this context.

The FS/wave plane analysis was first proposed in 1D \cite{YangLee1974} and further extended to higher dimensions in subsequent works \citep{kushwaha1993,Vasseur_1994,Jiao_1994}. The method is based on expressing the periodic displacement $U(x)$ as a Fourier series expansion that corresponds to a superposition of plane waves
\begin{equation*}
u(x) \approx \hat{U}_0 +  \hat{U}_1 \mathrm{e}^{\mathrm{i}{\xi_1}}x + ... +   \hat{U}_n \mathrm{e}^{\mathrm{i}{\xi_n}}x.\end{equation*}
The geometry of the periodic cell (determined by the density field, $\rho(x)$, and the stiffness field, 
$\mathbb{C}(x)$)
is also considered as a Fourier expansion, but in order to represent it accurately, the method considers only simple cases (cells made of spheres, cubes...) for which the series have an analytical expression. Then, the Fourier series representing the displacement is truncated to $n$ terms in order to limit the dimension of the problem, and introduced in the wave equation to derive a discrete system of equations of dimension $n$. The resulting linear system is dense and the eigenvalues of the system can be solved using a direct eigenvalue algorithm if the dimension $n$ is small. This approach has been used to obtain the dispersion diagram and band-gaps in layered materials \citep{YangLee1974} and composites \citep{kushwaha1993,Vasseur_1994}. The main limitations of this approach are, first, that it is restricted to simple geometries for which the Fourier series associated with the geometry has an analytical expression. The second limitation is that the approach results in a dense matrix in the real space which storage and eigenvalue analysis is computationally very demanding, limiting the maximum number of frequencies that can be considered in the expansion and, consequently, the accuracy of the results. 

The MST method is a mathematical model used to describe the propagation of a wave through a collection of scatterers, originally motivated by Lord Rayleigh in 1892 \cite{KORRINGA1994341}. When this formalism is applied to Bloch waves, the resulting approach allows determining the dispersion diagram of periodic media. The method, although developed for other wave equations like Schroedinger equation, can be used for phononic crystals with simple cell geometries \cite{waterman1976matrix}. In this case, the MST method present some advantages with respect to FS, since it has a much better convergence in the case of large or infinite contrast in elastic properties, as in the case of materials with empty space, like foams or architectural metamaterials. This approach has been applied to media with spherical inclusions \cite{Liu2000,Psarobas2000}, and to two-dimensional materials containing a solid matrix and fluid inclusions \cite{QIU2005765}. Although the MST method is advantageous with respect to FS for infinite mechanical contrast, both approaches share the same main limitation, requiring simple geometries of the periodic cell.

FEM has been profusely used to study acoustic wave dispersion in periodic heterogeneous solids, specially by the engineering community. The main and most clear benefit of FEM with respect to FS or MST is its ability in handling complex geometrical domains. The first FEM approach for elastic wave propagation in periodic media was proposed by \cite{mead1973general}, but only practically implemented in 1D, due to computational limitations. Since then, different approaches have been developed, which used FEM for studying dispersion in periodic media. As reported in \cite{palermo2020601}, two different FEM-based approaches have been considered. One is the Bloch Operator Finite Element Method \citep{Hussein2009}, which consists in introducing the Bloch formalism in the weak form of the wave equation and solving the eigenvalue problem using FEM. 
This method is usually combined with the approximation of the displacement functions in a reduced Bloch basis \citep{Hussein2009,palermo2020601}, 
reduced Bloch mode expansion, 
providing a very fast way to obtain the band diagram. However, an accurate reduced Bloch expansion relies in a clever selection of the reduced basis which is not trivial in the case of low-symmetry cells. The other possibility is the use of standard Finite Elements method, but imposing the wave condition through the boundary conditions. This approach is sometimes called Wave Finite Elements method. This alternative was first proposed in \cite{ZHONG1995485} and \cite{aberg_gudmunson1997} using a commercial FEM code by considering two identical meshes to split the complex-valued fields into real and imaginary parts, and applied to obtain the acoustic response of long-fibre composites. Many studies using Wave Finite Elements have been conducted to study the response of lattice-based metamaterials \cite{phani2006,KRUSHYNSKA201730} due to the growing interest in developing materials with special acoustic behavior. It must be noted that extracting the eigenvalues of the resulting large matrices is very expensive and many problems can only be efficiently resolved with the acceleration of the method using some type of reduced-order model technique \cite{BOUKADIA20181,palermo2020601}.


In the last 30 years, a class of numerical techniques alternative to FEM for solving periodic homogenization problems based on the Fast Fourier Transform (FFT) algorithm \cite{MS94} have become very popular. The original method was introduced by Moulinec and Suquet \citep{MS94,MS98} and is based in solving the equilibrium equation of a heterogeneous medium by transforming the heterogeneities into eigenstrains in a homogeneous reference elastic material. The solution of this equivalent problem can be expressed using Green's functions, which can be computed in Fourier space, giving rise to a Lippmann-Schwinger equation that can be solved iteratively. The main advantages of FFT-based approaches are their very efficient numerical performance (computational cost only grows a $n\log n$, where $n$ is the number of discretization points) and the possibility of using direct input from images of the material, avoiding complex meshing procedures. In recent years, several modifications of the original FFT-based approach have been proposed to improve the convergence of the iterative method for high mechanical contrast \citep{EYRE1999,MMS00,MONCHIET2013276, zeman2010accelerating,BD10,BRISARD2012,Kabel2014,KABEL2016,wichtanderson2021}. Also, alternative FFT-based formulations have been developed as Fourier-Galerkin approaches \citep{Vondrejc2014,Geers2016,Geers2017,Lucarini2019a,Lucarini2019b}, or related methods based on the use of displacements instead of strains as primary unknown field \citep{SOK2016,Lucarini2019c}. Although there are potential advantages (i.e. numerical efficiency, natural consideration of periodicity) for using FFT-based homogenization approaches to derive and solve the eigenvalue problem of elastic harmonic waves, this kind of applications is almost unexplored. The adaptation and use of FFT-based methods to solve Bloch wave problems appear as a very advantageous approach, specially in the case of very large and complex three-dimensional periodic cells with moderate mechanical contrast. This type of configurations are precisely the target of this work.

Based on the above considerations, the objective of this paper is twofold. On the one hand, we develop a new robust and efficient method based on FFT homogenization to obtain dispersion diagrams for arbitrary microstructures. The method overcomes some of the limitations of the existing approaches, and is specially efficient when applied to very large cells and moderate mechanical contrast. The periodic cell is discretized using a regular grid and, taking advantage of the simple expressions to calculate derivatives in Fourier space, we obtain a solution of the eigenvalues problem in Fourier space, defined by linear operators that do not require to form stiffness and mass matrices. 

On the other hand, the developed approach is used to study the acoustic response of elastic polycrystals, which is an ideal system to demonstrate the unique capabilities of the proposed method. Large three-dimensional unit cells containing hundreds of grains and near to 10$^6$ degrees of freedom are considered to determine dispersion relations in this type of materials, exploring in particular the effect of the texture and crystal anisotropy.

The paper is organized as follows. First, the Bloch wave formalism and the resulting differential equations in an elastic medium are reviewed, for completeness. In section 3, the proposed FFT-based methodology is presented. The method is then validated in section 4 using analytical results of simple cases and available numerical data in the literature. Section 5 presents applications of the method to polycrystals. Finally, section 6 provides conclusions and future work.

\section{Elastic waves in periodic solids: Bloch-Floquet theorem}
\subsection{One dimensional case}
The behavior of elastic waves in periodic solids can be mathematically described using the Bloch-Floquet formalism \citep{Bloch1929}, originally developed for describing the quantum wave functions of electrons in a crystal lattice. In one dimension, propagation is studied in an infinitely long material formed by the periodic repetition of unit cells with period $L$. The spatial position of a material point is defined by $x$ .The unit cell then consists in a 1D domain $\Omega$ of length $L$,  defined by the points $x\in[0,L]$. This unit cell can be made of different elastic materials, such that the microstructure can be defined using the functions $E(x)$ and $\rho(x)$, corresponding to the elastic stiffness and density respectively, which fulfills periodicity, i.e. $E(x)=E(x+nL)$, for $n\in \mathbb{Z}$. Let  $u(x,t)$ be a function defining the displacement of each point in space and time. Note that $u$ may correspond to either longitudinal or transversal displacements. From the Bloch-Floquet theorem, a wave propagating in the space can be described as a Bloch wave,
\begin{equation}\label{eq:Bloch_wave}
u(x,t)=U(x)\mathrm{e}^{\mathrm{i}kx-\mathrm{i} \omega t}
\end{equation}
where $k$ is the Bloch wave number that defines the spatial periodicity of the propagating wave, $\omega$ is the angular frequency at which the displacement of the material points oscillate and $U(x)$ is a periodic function. Note that, due to periodicity, \rev{only the values of $k\in[-\frac{\pi}{L},\frac{\pi}{L}]$ need to be analyzed, which correspond to the irreductible first Brioullin zone,\cite{Kittel_solidstate}}

The consequence of having Bloch waves propagating in the medium is that, for a wave with a given $k$, there are only some possible frequencies at which the wave can propagate in the periodic media. To find those frequencies and their corresponding shape ---modulated by the function $U(x)$ in the unit cell--- the conservation of linear momentum (eq. \ref{eq:waves1D}) is applied to a Bloch wave (eq. \ref{eq:Bloch_wave})
\begin{equation}
\label{eq:waves1D}
\frac{\mathrm{d}}{\mathrm{d}x}\left[ E(x)\frac{\mathrm{d}u(x,t)}{\mathrm{d}x}\right]-\rho(x)\frac{\mathrm{d}^2u(x,t)}{\mathrm{d}t^2}=0.
\end{equation}


The result is the following Helmholtz equation, defined  in the region $x\in[0,L]$, where the time dependency disappears due to the harmonic oscillation of the waves
\begin{eqnarray}
\label{eq:eigen_eq_1D}
-\left(\frac{\mathrm{d}}{\mathrm{d}x}+\mathrm{i}k\right)\left[ E(x)(\frac{\mathrm{d}}{\mathrm{d}x}+\mathrm{i}k)\right] \{U(x)\} = \omega^2 \rho(x)\{U(x)\} \nonumber \\
\mathcal{A}_{k}\{U(x)\} = \omega^2 \mathcal{M}\{U(x)\}\nonumber \\
\text{with} \ U(x=0)=U(x=L).
\end{eqnarray}
\rev{Equation (\ref{eq:eigen_eq_1D}) defines an eigenvalue problem, for any given $k$ non-trivial solutions of $U(x)$ exist only for certain values of $\omega^2$, corresponding to the eigenvalues of the equation. In eq.(\ref{eq:eigen_eq_1D}) $\mathcal{A}_k, \mathcal{M}$ are stiffness and mass differential operators and the problem consists in finding, for a given value of $k$, the eigenvalues $\omega^2$ (admisible frequencies) and their corresponding eigenvector functions $U(x)$.}

In the trivial case of a homogeneous medium ($E(x)=E$ and $\rho(x)=\rho$), after grouping terms, equation (\ref{eq:eigen_eq_1D})  results in
\begin{eqnarray}
E U''(x)+\mathrm{i} 2 E k U'(x)+(\rho\omega^2-Ek^2 )U(x)=0
\end{eqnarray}
In this case, the eigenvalues can be found analytically testing spatially harmonic solutions $U(x)=U_0\mathrm{e}^{\mathrm{i}\xi_n x}$ with periodicity  $\xi_n=n\frac{2\pi}{L}$, and $n\in \mathbb{Z}$. If these solutions are introduced in eq. (\ref{eq:eigen_eq_1D}), the resulting expression is
\begin{equation}
\label{disp_homo}
(\xi_n+k)^2=\frac{\omega_n^2\rho}{E}\rightarrow \omega_n = (\xi_n+k) \sqrt{ \frac{E}{\rho}} = (\xi_n+k) c\ , n=0,1,2...
\end{equation} 
where for each given $k$ there are infinite eigenvalues $\omega_n^2$. For each eigenvalue, the corresponding eigenvector is the plane wave $U(x)=U_0\mathrm{e}^{\mathrm{i}\xi_n x}$. The value $c=\sqrt{E/\rho}$ is the speed of sound in the homogenous medium. The displacement solution corresponding to this frequency is 
\begin{equation}
\label{eq:sol1D}
u(x,t)=U_0\mathrm{e}^{ \mathrm{i}(\xi_n+k) x - \mathrm{i}c (\xi_n+k) t} 
\end{equation}
that indeed is the general solution of a plane wave in a 1D medium. Note that in this case the relation between $k$ and $\omega$ is linear, and therefore the medium is non-dispersive.

\subsection{Waves in two and three dimensions}
\begin{figure}

\begin{center}
\includegraphics[width=.7\textwidth,valign=c]{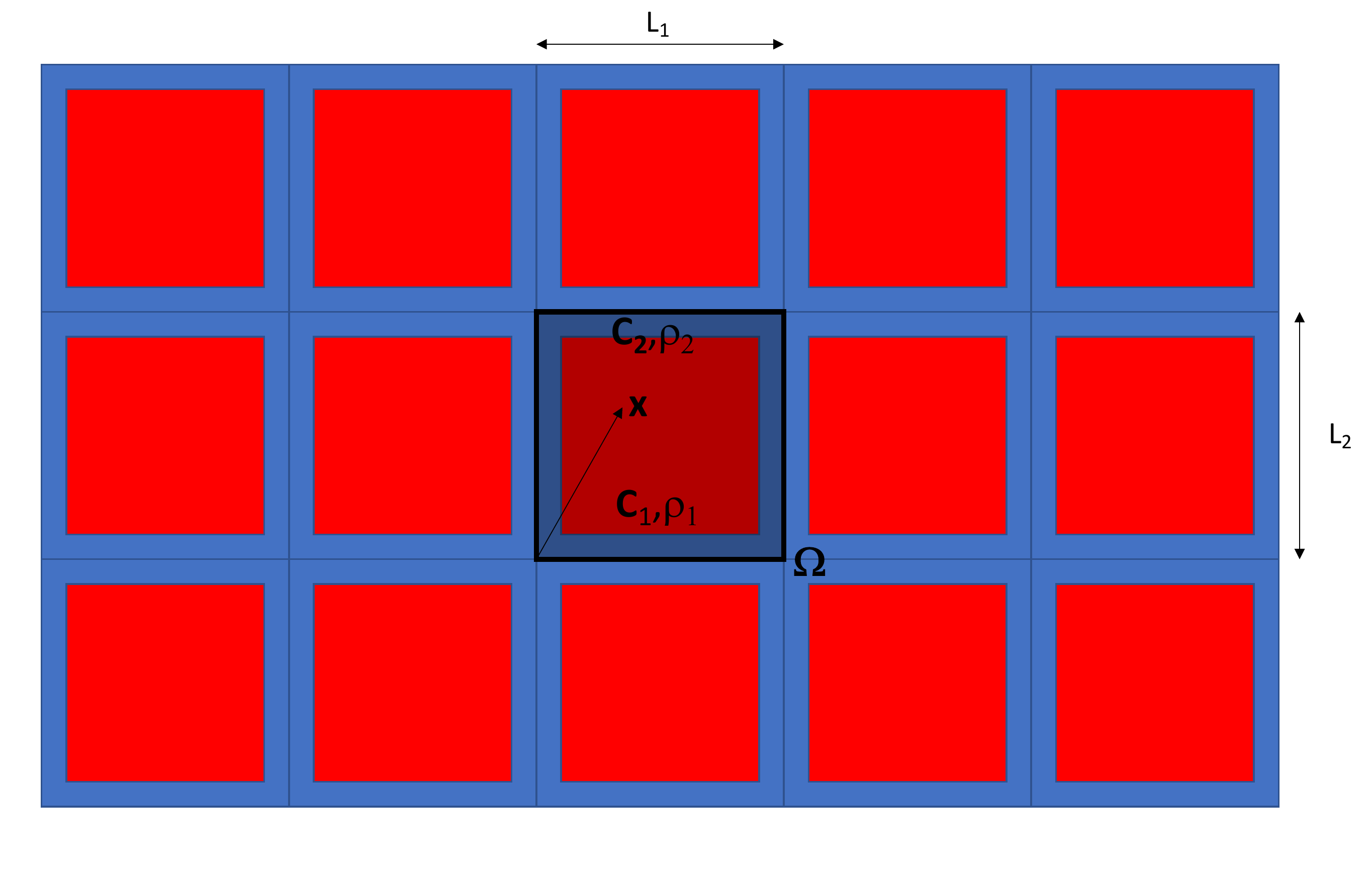}
\includegraphics[width=.29\textwidth,valign=c]{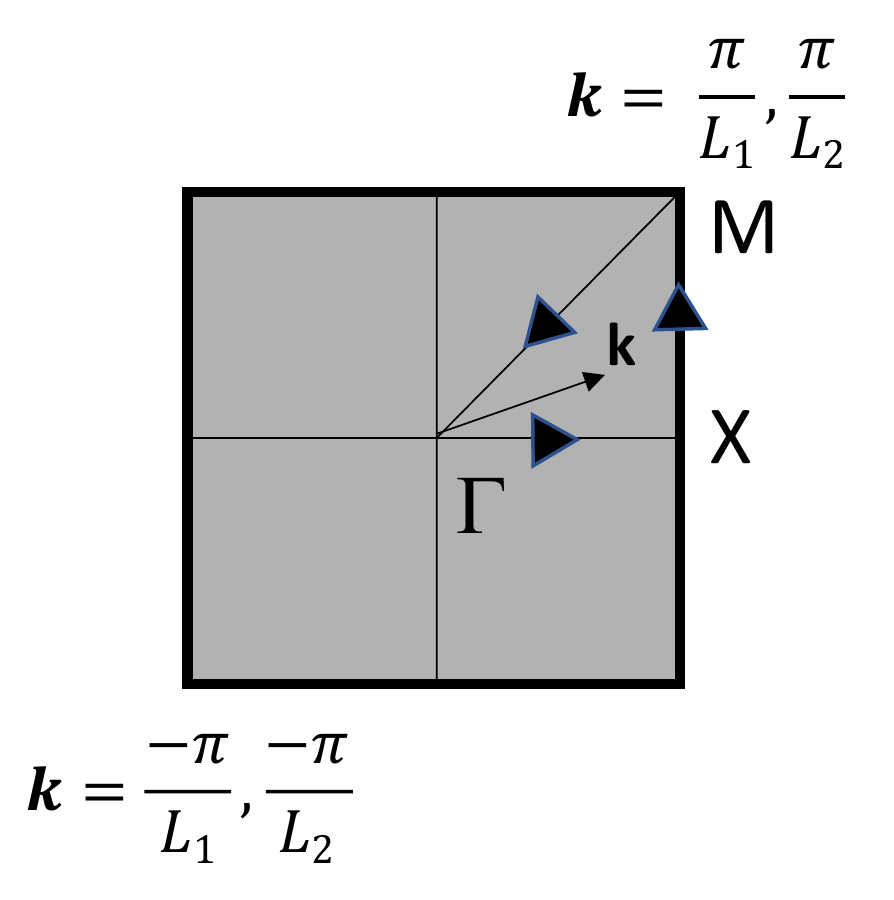}
\end{center}
\caption{Left: 2D periodic medium formed by a square array of square inclusions (red) embedded in a matrix (blue) with the unit cell $\Omega$ in grey. Right: The corresponding reciprocal cell including the definition of the admisible wave vectors.}\label{2D_periodic}
\end{figure}

In the case of waves propagating in a periodic 2D or 3D medium the problem is equivalent: for a wave with a given wave vector $\mathbf{k}$, obtain the possible values of frequencies $\omega$ for which that wave can propagate. 
The periodic medium in this case can be reduced to a unit cell, which for cubic periodicity corresponds to $\Omega=[0,L_1]\times [0,L_2]$ in 2D and  $\Omega=[0,L_1]\times[0,L_2]\times [0,L_3]$  in 3D. The cell is formed by different elastic phases and its microstructure is defined by the value of the elastic stiffness and density in each point, $\mathbb{C}(\mathbf{x})$ and $\rho(\mathbf{x})$, with these functions being periodic in the space, see Fig. \ref{2D_periodic}. The wave vectors $\mathbf{k}$ can be studied in a reduced domain due to the periodicity, $\mathbf{k}\in[0,\frac{\pi}{L_1}]\times [0,\frac{\pi}{L_2}]$, Fig. \ref{2D_periodic}.
 
The Bloch theorem implies that the displacement field of any material point is given by
\begin{equation}
\label{eq:plane_wave_3D}
\mathbf{u}(\mathbf{x},t)=\mathbf{U}(\mathbf{x})\mathrm{e}^{\mathrm{i}(\mathbf{k\cdot x}-\omega t)}
\end{equation}
where $\mathbf{U}(\mathbf{x})$ represents a periodic displacement vector field, identical for each unit cell, and $\mathbf{k}$ is a wave vector that represents the spacial periodicity of the propagating wave. The stress in the material is obtained from the displacement field of the Bloch wave using small strain theory,
\begin{equation}
\label{eq:plane_wave_stress_3D}
\boldsymbol{\sigma}(\mathbf{x},t)=\mathbb{C}(\mathbf{x}):\nabla^s \mathbf{u}(\mathbf{x},t)=\mathbb{C}(\mathbf{x}):(\nabla^s \mathbf{U}+\mathrm{i}\mathbf{U}\otimes \mathbf{k})\mathrm{e}^{\mathrm{i}(\mathbf{kx}-\omega t)}.
\end{equation}
where $\nabla^s$ is the symmetric gradient. Applying the conservation of linear momentum, $$
\nabla \cdot \boldsymbol{\sigma}(\mathbf{x},t)=\rho(\mathbf{x})\frac{\partial^2  \mathbf{u}(\mathbf{x},t)}{\partial t^2},$$
to the Bloch wave, and simplifying the resulting equation leads to 
\begin{eqnarray}
\label{eq:final}
\nabla \cdot \left[ \mathbb{C}(\mathbf{x}):(\nabla \mathbf{U}+\mathrm{i}\mathbf{U}\otimes \mathbf{k} )\right] +\mathrm{i} \mathbb{C}(\mathbf{x}):(\nabla \mathbf{U}+\mathrm{i} \mathbf{U}\otimes \mathbf{k})\cdot \mathbf{k}+
\rho(\mathbf{x})\omega^2\mathbf{U} = \mathbf{0} \
\end{eqnarray}
where the minor symmetry of the stiffness tensor was used. In components, the equation is written:
\begin{eqnarray}
\label{eq:final_components}
\left[ C_{ijkl}(\mathbf{x}):(U_{k,l}+\mathrm{i}U_{k}k_l) \right]_{,j}+\mathrm{i} C_{ijkl}(\mathbf{x}):(U_{k,l}+\mathrm{i}U_{k}k_l)k_j+\rho(\mathbf{x})\omega^2U_i=0
\end{eqnarray}

This equilibrium condition can be reorganized and written as the application of two linear differential operators to the vector function $\mathbf{U}$, 
\begin{eqnarray}
\mathcal{A}_{\mathbf{k}}\left\{ \mathbf{U(x)}\right\}=\omega^2\mathcal{M}\left\{\mathbf{U(x)}\right\}
\label{eq:eigen4}
\end{eqnarray}
where the two differential operators acting on $\mathbf{U(x)}$, $\mathcal{A}_{\mathbf{k}}\{\cdot\}$ and $\mathcal{M}\{\cdot\}$, are given by
\begin{eqnarray}
\label{eq:Areal3D}
\mathcal{A}_{\mathbf{k}}\left\{ \mathbf{U} \right\}=-\nabla \cdot \left[ \mathbb{C}(\mathbf{x}):(\nabla \mathbf{U}+\mathrm{i}\mathbf{U}\otimes \mathbf{k} )\right] +\mathrm{i} \mathbb{C}(\mathbf{x}):(\nabla \mathbf{U}+\mathrm{i} \mathbf{U}\otimes \mathbf{k})\cdot \mathbf{k}\\
\label{eq:Mreal3D}
\mathcal{M}\left\{\mathbf{U}\right\}=\rho(\mathbf{x})\left\{\mathbf{U}\right\}
\end{eqnarray}
The existence of non-trivial solutions of eq. (\ref{eq:eigen4}) for $\mathbf{U}$  implies solving a generalized eigenvalue problem, i.e. finding the eigenvalues $\omega$ and the eigenvector fields $\mathbf{U(x)}$ corresponding to each eigenvalue.

\section{FFT-based algorithm for acoustic dispersion diagrams}
In this section, the method to obtain dispersion diagrams of periodic heterogeneous media is presented. This method is based on a regular discretization of the unit cell representing the heterogeneous medium, and the transformation of the differential eigenvalue problem in the real space (eq. \ref{eq:eigen4}) into a discrete eigenvalue problem in Fourier space. The eigenvalue problem will be solved numerically using the efficient FFT algorithm and two different approaches, the restarted Lanczos method and the subspace iteration approach.

\subsection{Discrete eigenvalue problem in Fourier space in one dimension}  

Equation (\ref{eq:eigen_eq_1D}) can be transformed to Fourier space. For this, the derivatives therein can be computed using the definition of derivative in Fourier space,
\begin{equation}\label{eq:derivative}
\mathcal{F}(U'(x))=\mathrm{i}  \xi \mathcal{F}(U(x)) = \mathrm{i} \xi \hat{U}(\xi) .
 \end{equation}
 where $\mathcal{F}$ corresponds to the Fourier transform and $\hat{U}=\mathcal{F}(U)$.  After simplification, and expressing the equation as function of  $\hat{U}$, the resulting expression reads 
 \begin{equation}
 \label{eq:after_reference}
 (\xi +k) \mathcal{F} \left \{ E(x)\mathcal{F}^{-1} [\left( \xi+k \right)\hat{U} ] \right \}=\omega^2  \mathcal{F}(\rho(x)\mathcal{F}^{-1}(\hat{U})).
 \end{equation}
where $\mathcal{F}^{-1}$ is the inverse Fourier transform. It can be observed that the resulting equation is also an eigenvalue problem whose eigenvalues, $\omega^2$, are the same than the eigenvalues of the original partial differential equation. 

The left-hand side of eq. (\ref{eq:after_reference}) is the Fourier transform of the linear operator $\mathcal{A}_k$, and acts on displacement fields defined in Fourier space
 \begin{eqnarray}
  \label{eq:defAk}
 \hat{\mathcal{A}}_k\left\{\hat{U}\right\}:= (\xi+k)\mathcal{F}\left\{E(x)\mathcal{F}^{-1}[\left( \xi+k \right)\hat{U}] \right\}.
 \end{eqnarray}
This new operator is also linear in $\hat{U}$, since it just contains products with other functions and direct and inverse Fourier transformations. The right-hand side of eq. (\ref{eq:after_reference}) contains the Fourier transform of the mass linear operator.
  \begin{eqnarray}
   \label{eq:defM}
 \hat{\mathcal{M}}\left\{\hat{U}\right\}:= \mathcal{F}(\rho(x)\mathcal{F}^{-1}(\hat{U})).
 \end{eqnarray}
The eigenvalue equation can be written in Fourier space as
 \begin{eqnarray}
\hat{\mathcal{A}}_k(\hat{U}(\xi))=\omega^2\hat{\mathcal{M}}(\hat{U}(\xi))
\label{eq:eigen_1Dlin}
\end{eqnarray}
which corresponds to a generalized eigenvalue equation for the linear operator $\hat{\mathcal{A}}_k$ (stiffness) and the mass linear operator $\hat{\mathcal{M}}$. The linear operators are Hermitian, as  the original differential operators, and therefore the eigenvalues are positive real numbers defining the possible values of the angular frequency $\omega^2$. The eigenvectors, $\hat{U}(\xi)$, are complex functions representing  the mode corresponding to that frequency in the Fourier space.

In order to numerically solve the problem, the real space will be discretized into a partition with $N$ equal parts, and the values of the functions will be represented by their value in the center of each part,
\begin{eqnarray}
E(x)  \rightarrow E(x_n)=E_n,\nonumber \\
\rho(x)\rightarrow \rho(x_n)= \rho_n \nonumber \\
U(x) \rightarrow U(x_n)=U_n
\end{eqnarray}
with  $x_n=\frac{L}{N}(\frac{1}{2} + n)$ and $n\in[0,N-1]$. The Fourier transform and its inverse will correspond then to the discrete Fourier transform that can be computed with the efficient FFT algorithm. The corresponding $N$ discrete frequencies in the Fourier space are 
\begin{equation}\label{frequencies}
\xi_k=\left\{ \begin{array}{c} 
\frac{2\pi}{L} (k-\frac{(N-1)}{2}) \quad \text{ for N odd}\\
\frac{2\pi}{L} (k-\frac{N}{2}) \quad \text{ for N even}
\end{array}
\right.  \text{ for } \quad k =0, \dots, N-1 
\end{equation}

The discrete version of the eigenvalue problem  in eq. (\ref{eq:eigen_1Dlin}) will be solved numerically in Fourier space. To this aim, the linear operators defined in eqs. (\ref{eq:defAk},\ref{eq:defM}) will be directly used in an iterative algorithm, avoiding to form the corresponding matrices. 

\subsection{Discrete eigenvalue problem in Fourier space in 2 and 3 dimensions}  

The eigenvalue differential problem (\ref{eq:eigen4}) will be transformed to Fourier space using the definition of differential operators in Fourier space. After simplification, and making use of the stiffness symmetries, the linear operators given in the real space by eqs. (\ref{eq:Areal3D},\ref{eq:Mreal3D}), have simple expressions in Fourier space, given by
 \begin{eqnarray}
\label{eq:hatA}
\hat{\mathcal{A}}_{\mathbf{k}}\left\{\hat{\mathbf{U}}(\boldsymbol{\xi})\right\}:=\mathcal{F}\left\{ \mathbb{C}(\mathbf{x}):\mathcal{F}^{-1}(\boldsymbol{\xi}+\mathbf{k})\otimes\hat{\mathbf{U}} ) \right\} \cdot (\boldsymbol{\xi}+\mathbf{k})  
\end{eqnarray} 
or in components,
 $$
\hat{\mathcal{A}}_{\mathbf{k}i}\left\{\hat{\mathbf{U}}\right\}:=(\xi_j+k_j) \mathcal{F} \left\{ C_{ijkl}(\mathbf{x})\mathcal{F}^{-1}((\xi_l+k_l)\hat{U}_k )\right\} $$
and
\begin{equation}
\label{eq:hatM}
\hat{\mathcal{M}}\left\{\hat{\mathbf{U}}(\boldsymbol{\xi}) \right\}:= \mathcal{F}\left\{\rho(\mathbf{x})\mathcal{F}^{-1}(\mathbf{\hat{U}})\right\}
\end{equation}
being both linear operators Hermitian.

The generalized eigenvalue problem in Fourier space corresponds to
\begin{equation}
\label{eq:eigen_fourier_3D}
\hat{\mathcal{A}}_{\mathbf{k}}\left\{\hat{\mathbf{U}}(\boldsymbol{\xi})\right\}=\omega^2\mathcal{M}\left\{\hat{\mathbf{U}}(\boldsymbol{\xi})\right\}
\end{equation}
where the eigenvalues are real and positive and define $\omega^2$ (equal to the angular frequency in the eigenvalue problem in real space). For each eigenvalue $\omega^2$, $\hat{\mathbf{U}}$ is its corresponding eigenvector, a complex function defined in Fourier space.

In order to solve the most general 3D problem numerically, the domain is discretized in $N_1\cdot N_2\cdot N_3$ equispaced voxels, with centers given by
\begin{equation}
\mathbf{x}_{i,j,k}=\left(\frac{L_1}{N_1}(\frac{1}{2}+i),\frac{L_2}{N_2}(\frac{1}{2}+j),\frac{L_3}{N_3} (\frac{1}{2}+k)   \right)
\end{equation}
with $(i,j,k) \in [0,N_1-1]\times[0,N_2-1]\times[0,N_3-1]$. The value of the material properties in each voxel, $\mathbb{C}(\mathbf{x}_{i,j,k})$ and $\rho(\mathbf{x}_{i,j,k})$ correspond to the property of the material occupying that position. The displacement vector is discretized using its value at the center of each voxel, $\mathbf{u}(\mathbf{x}_{i,j,k})$.
 

The discrete fields in Fourier space have the same dimension as their counterparts in real space. Both Fourier and inverse Fourier discrete transforms are computed using the FFT algorithm. The corresponding $N_1\cdot N_2 \cdot N_3$ discrete frequencies in the Fourier space are   \rev{
\begin{equation}
\boldsymbol{\xi}_{k_1,k_2,k_3}=\left( {\xi}_{k_1}, {\xi}_{k_2}, {\xi}_{k_3}\right)= \left (k_1\frac{2\pi}{L_1},k_2\frac{2\pi}{L_2},k_3\frac{2\pi}{L_3}\right )
\end{equation}
 with $\xi_{k_1},\xi_{k_2},\xi_{k_3}$ defined as in the 1D case, (eq. \ref{frequencies}).}

The discrete eigenvalue problem in Fourier space, given by eqs. (\ref{eq:eigen_1Dlin}) and (\ref{eq:eigen_fourier_3D}) for one and higher dimensions, respectively, will be solved numerically using two alternative methods.

\subsection{Eigenvalue extraction}
The most critical aspect of the proposed method is an efficient extraction of the smallest eigenvalues of the discrete eigenvalue problem, defined by eq. (\ref{eq:eigen_1Dlin}) for 1D and by eq. (\ref{eq:eigen_fourier_3D}) for 3D. Two iterative methods are used to find the smallest eigenvalues in the discrete Fourier space, the implicitly restarted Lanczos method \cite{lanczos1950, Lehoucq2001} and the subspace iteration method \cite{bathe1980accelerated,bathe2013subspace}. Both methods can be adapted to use discrete linear operators and therefore do not \rev{require} to form a matrix. The details of the implementation and use of each method are described below. A detailed comparison between the two approaches can be found in \cite{lanczos_vs_subspace}.

\subsubsection*{Implicitly Restarted Lanczos Method}\label{lanczos}

The Arnoldi method \cite{arnoldi1951} is an iterative algorithm used to extract the largest eigenvalues (and their corresponding eigenvectors) of general matrices by constructing an orthonormal basis of the Krylov subspace. In the present particular case of Hermitian matrices, the Arnoldi method is known as the Lanczos method, proposed by C. Lanzcos in \cite{lanczos1950}. In this method, discrete linear operators can be used without an explicit assembly of their corresponding matrix, in our case, directly using the actuation of the discrete linear operators  $\hat{\mathcal{A}}_k$ and $\hat{\mathcal{M}}$ on displacement vectors expressed in Fourier space $\hat{\mathbf{U}}$.   Here we use the implicit restarted Arnoldi approach \cite{Lehoucq2001} (IRAM) and its implementation in the ARPACK package \cite{ARPACK}. This approach combines the Lanczos factorization with an implicitly shifted QR scheme, aiming at splitting the truncated matrix in the Krylov subspace as a product of an orthogonal and an upper triangular matrices. 

The resulting algorithm allows to efficiently compute a set of the largest eigenvalues of very large operators. If a specific eigenvalue spectrum is desired, e.g. the smallest eigenvalues, the shift theorem has to be applied. 
The original problem consists in  finding the set of the $p$ smallest eigenvalues from a total of $n$, $\lambda_1<\lambda_2<...<\lambda_p<...\lambda_n$, 
\begin{equation}
\lambda_p ,\  \hat{\mathbf{U}}_p \quad \mathrm{such}\  \hat{\mathcal{A}}_k\left\{\hat{\mathbf{U}}_p\right\}= \lambda_p\hat{\mathcal{M}}\left\{\hat{\mathbf{U}}_p\right\}
\label{eq:noshift}
\end{equation}
which, using the shift theorem for eigenvalues near 0 (smallest ones), can be converted to find a the set of the $p$ largest eigenvalues $\beta$, $\beta_1>\beta_2>...\beta_p>...\beta_n$
\begin{equation}
 \beta_p ,\  \hat{\mathbf{U}}_i \quad \mathrm{such}\  \hat{\mathcal{A}}^{-1}_k\left\{\hat{\mathbf{U}}_p\right\}= \beta_p\hat{\mathcal{M}}^{-1}\left\{\hat{\mathbf{U}}_p\right\}
 \label{eq:shift}
\end{equation}
where $\lambda_p=1/\beta_p,p=1,...,n$. $\hat{\mathcal{A}}^{-1}_k$ is the inverse operator, defined as
 \begin{eqnarray}
 \hat{\mathcal{A}}_k^{-1}(\hat{\mathbf{Y}}):=\hat{\mathbf{U}} \mid  \hat{\mathcal{A}}_k(\hat{\mathbf{U}}) = \hat{\mathbf{Y}}
 \label{eq:def_invop}
 \end{eqnarray}
and $\hat{\mathcal{M}}^{-1}$ is the inverse mass operator, with a close expression obtained from eq. (\ref{eq:hatM}) given by
 \begin{eqnarray}
 \hat{\mathcal{M}}^{-1}(\hat{\mathbf{Y}}):= \mathcal{F}\left\{\frac{1}{\rho(\mathbf{x})}\mathcal{F}^{-1}(\mathbf{\hat{U}})\right\}.
 \label{eq:def_invM}
 \end{eqnarray}
The eigenvector corresponding to the eigenvalue $\lambda_p$ in eq. (\ref{eq:noshift}) is the same than eigenvector corresponding to the eigenvalue $\beta_p=1/\lambda_p$ in eq. (\ref{eq:shift}). 

\subsubsection*{Subspace iteration method}
This method was proposed by K.J. Bathe in the late 70s \cite{bathe1980accelerated} in the context of the Finite Element method for full structures and using the assembled stiffness and mass matrices. The method is adapted here to solve the eigenvalue problem of eq. (\ref{eq:eigen_fourier_3D}), which unlike the original approach, is defined in a complex space and makes use of linear operators over complex vectors instead of using full real matrices. 
A short description of the method and its adaptation is presented. For more details about the original method, the reader is referred to the relevant literature \cite{bathe1980accelerated,bathe2013subspace}.

The aim of the method is to obtain the subset of the smallest eigenvalues $\lambda_p=\omega^2_p,p=1,...n$ and their corresponding eigenvectors in the Fourier space, $\hat{\mathbf{U}}_p,p=1,...,n$, for the eigenvalue problem defined in eq.( \ref{eq:eigen_fourier_3D}). Those pairs of eigenvalues and eigenvector fulfill
\begin{equation}
\hat{\mathcal{A}}_{\mathbf{k}}\left\{\hat{\mathbf{U}_p}(\boldsymbol{\xi})\right\}=\lambda_i\mathcal{M}\left\{\hat{\mathbf{U}_p}(\boldsymbol{\xi})\right\}, p=1,...,n\label{eq:subspace}
\end{equation}
with $\lambda_1<\lambda_2<...<\lambda_n$

The method is iterative. Let $q>n$ (it is suggested $q=\min \{2n, n + 8\}$) be the number of vectors of the subspace considered. The iterative procedure starts with an initial set of vectors, $\hat{\mathbf{U}}_p, p=1,...,q$. In the original approach \cite{bathe1980accelerated,bathe2013subspace}, guidance is given to obtain a good initial set. However, in this work we follow a different strategy. If the eigenvalue problem has been solved for a vector number $\mathbf{k_0}$ near $\mathbf{k}$ (which is the case when obtaining a full dispersion diagram), the eigenvectors of that problem are used as initial vectors of the current computation. In the absence of a close solution, the initial eigenvectors are taken as unit vectors corresponding to the lowest frequencies. 

Let $s$ be the iteration number, the subspace iteration method algorithm applied to obtain the eigenvalues and eigenvectors  of eq. (\ref{eq:eigen_fourier_3D}) is given in Algorithm \ref{alg:subspace}. The main modification respect the original approach is that the linear equations 
\begin{equation}
[\hat{\mathcal{A}}_{\mathbf{k}}]\mathbf{x}=\mathbf{b},\label{eq:inverse}
\end{equation}
which are solved in block for a set of $b$'s in the original formulation, are solved here sequentially and iteratively using the conjugate gradient method. This is expressed in Algorithm \ref{alg:subspace} as the inverse operator given in eq. (\ref{eq:def_invop})
where the details for an efficient evaluation are given in next section.

\RestyleAlgo{boxed}
\begin{algorithm}
\DontPrintSemicolon
{\bf{Data:}} Vector number $\mathbf{k}$,  $TOL_{eig}$\\ 
\ Operators $\hat{\mathcal{A}}_{\mathbf{k}}\{\cdot  \}$, $\hat{\mathcal{M}}\{\cdot \}$ and $\hat{\mathcal{A}}^{-1}_{\mathbf{k}}\{\cdot \}$ all defined in
$\mathbb{C}^n\rightarrow \mathbb{C}^n$\\
\ Initial vectors (iteration $s=0$, $\hat{\mathbf{U}}^{(s=0)}_p, p=1,...,q, \hat{\mathbf{U}}_p \in \mathbb{C}^n$, 

 \While{$err_1>TOL_{eig} \ \& err_2>10\ TOL_{eig}$ }
 {
$ \mathbf{v}_p = \hat{\mathcal{M}}\{\hat{\mathbf{U}}^{(s)}_p \}\ p=1,...,q $\\
$\bar{\mathbf{U}}(\vdots,p)=\hat{\mathcal{A}}^{-1}_{\mathbf{k}}\{  \mathbf{v}_p  \} \ p=1,...,q \ ; \bar{\mathbf{U}} \in  \mathbb{C}^{n\times q} \ ^{(1)}$\\
$\mathbf{Kn}(\vdots,p) =  \mathbf{v}_p , \ p=1,...,q \ ; \mathbf{Kn} \in  \mathbb{C}^{n\times q}$\\
$\mathbf{Mn}(\vdots,p) =  \hat{\mathcal{M}} \{ \bar{\mathbf{U} } ( \vdots,p)	\}  \ i=1,...,q   ;  \mathbf{Mn} \in \mathbb{C}^n \times \mathbb{C}^q $\\
$\mathbf{K} = \text{conj}(\bar{\mathbf{U}}^T)\cdot \mathbf{Kn}, \ \mathbf{K} \in \mathbb{C}^q\times \mathbb{C}^q$\\
$\mathbf{M} = \text{conj}(\bar{\mathbf{U}}^T)\cdot \mathbf{Mn}, \ \mathbf{M} \in \mathbb{C}^q\times \mathbb{C}^q $\\
$\lambda^{(s+1)}_p ,\mathbf{q}_p = \text{EIG}(\mathbf{K},\mathbf{M}),  p=1,...,q \ ^{(2)}$ \\
$ \hat{\mathbf{U}}^{(s+1)}_p = (\bar{\mathbf{U}}\cdot [\mathbf{q}_1 | \mathbf{q}_2 | ... | \mathbf{q}_q] )^T$\\
$err_1 = \left( \sum_{1}^{n} (\lambda^{(s+1)}_p-\lambda^{(s)}_p)^2 / \sum_{1}^{n} (\lambda^{(s)}_p)^2 \right)^{1/2}$\\
$err_2 = \max_{p\leq n} \left \{ \hat{\mathcal{A}}_{\mathbf{k}}\{ \hat{\mathbf{U}}^{(s+1)}_p \} -\lambda_p   \hat{\mathcal{M}} \{ \hat{\mathbf{U}}^{(s+1)}_p \} \right \}$ 
 }
{\bf Output:} $\lambda_p, \hat{\mathbf{U}}_p,  i=1,...,n$\\ 
$^{(1)}$ The inverse operator is computed numerically using Algorithm \ref{eq:inverse}\\
$^{(2)}$ Is a generalized eigenvalue problem of small matrices (dimension $q$) which can be easily solved with any method. Here the QR algorithm is used\\
\caption{Algorithm to find the eigenvalues and eigenvectors defined by the linear operators $\hat{\mathcal{A}}_k,\hat{\mathcal{M}}$ using the subspace iteration method.}   
\label{alg:subspace}
\end{algorithm}

\subsubsection*{Inverting the A operator and method comparison}

As presented above, both  the subspace iteration method and the Lanczos algorithm with shifting require the evaluation of  the inverse of the discrete operator $ \hat{\mathcal{A}}_k$. Since the operator is Hermitian, and all its eigenvalues are positive for $\mathbf{k}\neq\mathbf{0}$,  the operator is invertible and a unique solution of eq. (\ref{eq:inverse}) can be found. 

For one dimension the inverse operator $\mathcal{\hat{A}}_k^{-1}(\hat{Y})$ can be obtained explicitly after some manipulation, and corresponds to
 \begin{eqnarray}
 \hat{\mathcal{A}}_k^{-1}(\hat{Y}):=\frac{1}{\xi+k}\mathcal{F}\left\{\frac{1}{E(x)}\mathcal{F}^{-1}[\frac{1}{\xi+k}\hat{Y}] \right\}
 \end{eqnarray}

For higher dimensions, there is no explicit solution of the inverse operator, and a numerical procedure is needed to evaluate  eq. (\ref{eq:def_invop}). The method  needs to be very efficient, since the inverse operator is used many times in both eigenvalue algorithms. Ideally, a direct method would be optimal to solve the equation since, once factorized, each resolution of the system would require just the substitution. However, operator $ \hat{\mathcal{A}}_k$ written as a matrix is fully dense, so the storage grows as $\mathcal{O}(n^2)$ and the number of operations $\mathcal{O}(n^3)$ limits this approach to very small cases. The alternative is the use of a Krylov-based iterative solver, in which the coefficient matrix is not stored. Thanks to the positive definiteness of the operator, the conjugate gradient method arises as an ideal method to perform this inverse numerically. Nevertheless, the number of iterations needed for solving directly the linear system defined in eq. (\ref{eq:def_invop}) is very large, and the use of a preconditioner becomes essential. A preconditioner is an approximate operator of the inverse, which can be computed with small computational cost. If the unit cell is considered to be homogeneous, using $\mathbb{C}^0$ in the expression $\hat{\mathcal{A}}_k$, the resulting operator does not require to compute Fourier transforms
\begin{eqnarray}
\hat{\mathcal{A}}^{app}_{\mathbf{k},i}\left\{\hat{\mathbf{U}}\right\}:=
(\xi_j+k_j) \mathcal{F} \left\{ C^0_{ijkl}\mathcal{F}^{-1}((\xi_l+k_l)\hat{Y}_k )\right\} = \nonumber \\C^0_{ijkl} (\xi_j+k_j) (\xi_l+k_l) \hat{U}_k(\boldsymbol{\xi}).
\end{eqnarray}
The previous operator just corresponds to the product of the matrix $\mathbf{P}(\boldsymbol{\xi})$, given by the acoustic tensor,
\begin{equation}
P_{ik}=C^0_{ijkl} (\xi_j+k_j) (\xi_l+k_l)
\label{eq:matrixP}
\end{equation}
and the value of $\hat{\mathbf{U}}(\boldsymbol{\xi})$ for each Fourier voxel $\boldsymbol{\xi}$ and therefore the operator can be inverted directly as
\begin{eqnarray}
\hat{\mathcal{A}}^{app,-1}_{\mathbf{k}}\left\{\hat{\mathbf{Y}}\right\}=\hat{\mathcal{P}} \left\{\hat{\mathbf{Y}}\right\} = \mathbf{P}^{-1}(\boldsymbol{\xi}) \hat{\mathbf{Y}}(\boldsymbol{\xi})
\label{eq:precond}
\end{eqnarray}
being the $3\times 3$ matrix inverse computed only once, at the beginning of the algorithm. Note that the preconditioner $\hat{\mathcal{P}}$ is equivalent to the one derived in \cite{Lucarini2019c} for the displacement-based FFT approach, DBFFT, but replacing the Fourier frequency with the sum of that frequency with the Bloch wave vector , $\boldsymbol{\xi}\leftarrow \boldsymbol{\xi} +\mathbf{k} $. For the value of $\mathbb{C}^0$, the volume averaged stiffness provides good numerical performance, as shown in \cite{Lucarini2019c}  for static simulations. It must be remarked that the use of the preconditioner is essential for the application of the method since it reduces the number of iterations by orders of magnitude.

The resulting algorithm to compute the inverse operator is written in pseudo-code in Algorithm \ref{eq:inverse}.
\RestyleAlgo{boxed}
\begin{algorithm}

\SetKwFunction{FMain}{InvA}
\SetKwProg{Fn}{Function}{:}{}
\Fn{\FMain{$\mathcal{A},  \mathcal{P}, \mathbf{Y} , TOL_{lin} $} }{
Call to Conjugate Gradient solver for linear operator function $\mathcal{A}$, preconditioner function $\mathcal{P}$ and independent term $\mathbf{Y} $ being a vector field\\
    \smallskip
    {\bf{return}} $ \mathbf{U} \, \,   | \, \,  \mathcal{A} ( \, \mathbf{U}  \, )  = \mathbf{Y} \, $  and  $\| \mathcal{A} ( \,   \mathbf{U}  \, ) - \mathbf{Y}  \| < TOL \cdot \|   \mathbf{Y} \| $ 
}

\SetKwFunction{FMain}{$\mathcal{P}$}
\SetKwProg{Fn}{Function}{:}{}
\Fn{\FMain{$\hat{\mathbf{U}} (\boldsymbol{\xi})$}}
{
    Compute the preconditioner linear operator $\mathcal{P}\left \{  \cdot \right \}$, eqs. (\ref{eq:precond}, \ref{eq:matrixP})\\
    \smallskip
    {\bf{return}} $\mathcal{P}\left \{ \hat{\mathbf{U}}(\boldsymbol{\xi}) \right \}=P^{-1}_{ik} \hat{U}_k(\boldsymbol{\xi}) $ 
}

\caption{Function for the inverse of the stiffness linear operator }     
\label{eq:inverse}
\end{algorithm}

Concerning the numerical performance of each method, it depends on the particular system and application. In general, for obtaining the eigenvalues of a single wave vector $\mathbf{k}$ (not the full diagram), Lanczos tends to be faster since it does not require an initial guess of the eigenvalues to be solved. In the case of a full dispersion diagram, subspace iteration method may be competitive if the increments in $\mathbf{k}$ are small, since the previously converged eigenvectors can be used as a guess for the new  $\mathbf{k}$. Both methods were used in the validation  of the model (section 4) and it was observed that for these cases the Lanczos method performance was always superior. Therefore, the rest of the cases presented here were computed using the Lanczos method.

\subsection{Full Algorithm}
The final algorithm to compute the dispersion diagram is just the application of one of the eigenvalue methods presented for obtaining the eigenvalues of eq. (\ref{eq:eigen_fourier_3D}) for a set of wave vectors $\mathbf{k}$ which cover the area of interest. The algorithm is presented in Algorithm \ref{Alg:total}
\RestyleAlgo{boxed}
\begin{algorithm}
Obtain the first $n$ bands for a list of wave vectors $\mathbf{k}^r, r=1,...,N$ \\
 \KwData{Microstructure and properties:  $\mathbb{C}(\mathbf{x}),\rho(x)$,  linear operator inversion tolerance, $TOL_{lin}$, eigenvalue problem  tolerance $TOL_{eig} $} 
{\it (a) Discretization of unit cell}, storage of $\mathbb{C}(\mathbf{x}_{i,j,k})$, $\rho(\mathbf{x}_{i,j,k})$ and $\boldsymbol{\xi}_{i,j,k}$ \\
 {\it{(b) \underline{Loop in wave vectors} } } \\
\For { $r$ \text{in} $N$ } {
$\mathbf{k}=\mathbf{k}^r$\\
\If {method == Lanczos}
{$\lambda^r_p, \hat{\mathbf{U}}^r_p$ = Lanczos($\hat{\mathcal{A}}^{-1}_k,\hat{\mathcal{M}}, TOL_{lin},TOL_{eig}), p=1,...n $ (Section \ref{lanczos}) }

\If {method == Subspace} {
$\lambda^r_p, \hat{\mathbf{U}}^r_p $ = Subspace($\hat{\mathcal{A}}_k,\hat{\mathcal{A}}^{-1}_k,\hat{\mathcal{M}},\hat{\mathbf{U}}^{r-1}, TOL_{lin},TOL_{eig}) ,p=1,...,n$ (Algorithm \ref{alg:subspace})
}
$\omega^r_p=\sqrt{\lambda^r_p}, p=1,..,n $ \\
$\mathbf{U}^r_p = \mathcal{F}^{-1} \{ \hat{\mathbf{U}}^r_p  \}$\\

    \SetKwFunction{FMain}{$\hat{\mathcal{A}}_k$}
\SetKwProg{Fn}{Function}{:}{}
    \Fn{\FMain{ $\hat{\mathbf{U}} (\boldsymbol{\xi})$ } }
    {
    Compute the stiffness linear operator $\mathcal{\hat{A}}_k \left\{  \cdot \right\}$, eqs. (\ref{eq:hatA})\\
    \smallskip
    {\bf{return} $\hat{\mathcal{A}}_k(\hat{\mathbf{U}})$}   
    }

}
    \SetKwFunction{FMain}{$\hat{\mathcal{M} }$ } 
\SetKwProg{Fn}{Function}{:}{}
    \Fn{\FMain{$\hat{\mathbf{U}} (\boldsymbol{\xi})$}}
    {
    Compute the mass linear operator $ \mathcal{\hat{M}} \left\{  \cdot \right\}$, eqs. (\ref{eq:hatM})\\
    \smallskip
    {\bf{return} $\hat{\mathcal{M}}(\hat{\mathbf{U}})$ }   
    }
    \caption{Full algorithm for obtaining a dispersion diagram}
\label{Alg:total}
\end{algorithm}

The full FFT-based method for obtaining dispersion diagrams (Algorithm \ref{Alg:total}) has been programmed in Python within the FFTMAD code developed by S. Lucarini and J. Segurado \cite{Lucarini2019a}

\section{Validation of the method}
In this section, the proposed methodology will be validated using exact solutions when available (homogeneous material and laminate material). In more complex cases, where no explicit solution exists, the method is compared with published results from the literature, obtained using FS/plane wave analysis.

\subsection{Homogeneous solid}
First, the proposed framework will be used to obtain the dispersion relation for elastic waves traveling in a homogeneous material in one and two/three dimensions. In both cases, analytical solutions can be found for the eigenvalues $\omega$ as function of the wave vector $\mathbf{k}$. In one dimension, the different eigenvalues are given by equation (\ref{disp_homo}), where the different $n$ modes for a particular wave number are given by $\omega_n=n\frac{2\pi}{L}$. In the case of higher dimensions, considering the eigenvalue equation in Fourier space (eq. \ref{eq:eigen_fourier_3D}) constant values of $\mathbb{C} (\mathbf{x})=\mathbb{C}$ and $\rho(\mathbf{x})=\rho$ leads to
\begin{equation} \label{eq:anal_eigen_3D}
C_{ijkl}(\xi+k)_j(\xi+k)_l \hat{U}_k =\omega^2 \rho  \hat{U}
\end{equation}
where  $\boldsymbol{\xi} = \frac{2\pi}{L} [p,q,r]$ defines the different eigenvalues for a particular $\mathbf{k}$ being $p,q,r \in \mathbb{Z}$. Equation (\ref{eq:anal_eigen_3D}) corresponds to an eigenvalue problem of dimension 2 or 3, which can be solved analytically. If 
a matrix $\mathbf{P}$
(acoustic tensor) is defined for each $\boldsymbol{\xi}$
$$ P_{ik} = C_{ijkl}(\xi+k)_j(\xi+k)_l, $$
the eigenvalue problem to solve for each given $\boldsymbol{\xi}$ is
$$\frac{1}{\rho} \mathbf{P}\  \mathbf{U}  =\omega^2 \mathbf{U}$$ 
and the two/three eigenvalues (depending on the dimension of the problem) correspond to the longitudinal and transverse modes for that particular $\boldsymbol{\xi}$.

\begin{figure}

\includegraphics[width=.49\textwidth]{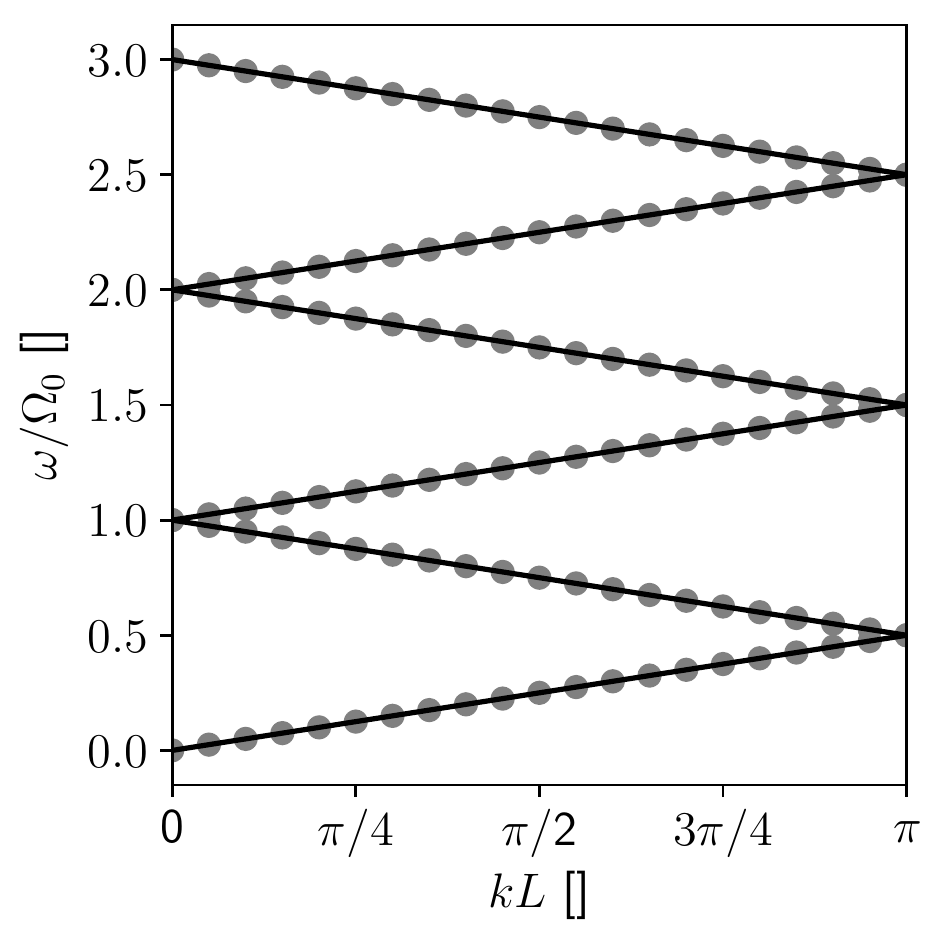}
\includegraphics[width=.49\textwidth]{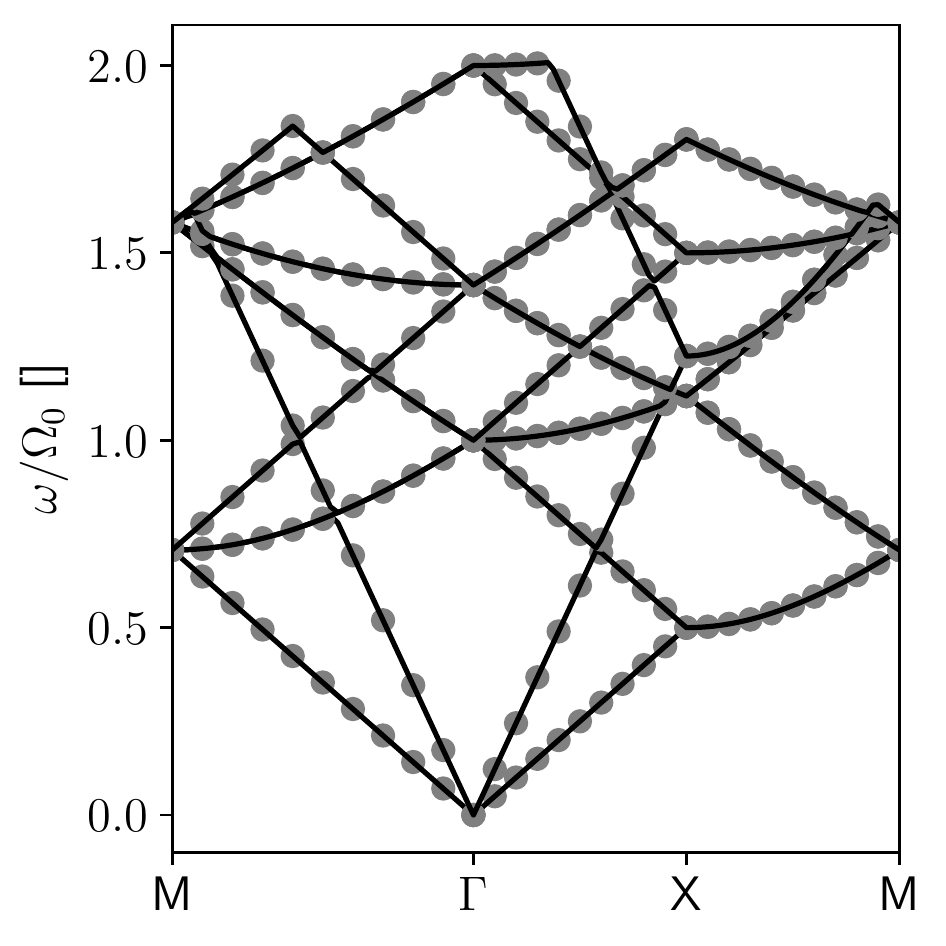}
\caption{Dispersion diagram for a one dimensional (left) and two dimensional (right) homogeneus materials}
\label{dis_hom}
\end{figure}

The homogeneous material simulated as example is an epoxy resin whose properties are given in table \ref{table:1}, taken from \cite{Vasseur_1994}. The  dispersion diagrams obtained numerically using the proposed FFT-based approach are given in Fig. \ref{dis_hom} for one and two dimensions, together with the analytical solution given by eqs. (\ref{eq:sol1D}) and (\ref{eq:anal_eigen_3D}) respectively. In the 1D case, $k$ values are chosen $0\leq k\leq \pi/L$, while in the two dimensional case, the vectors $\mathbf{k}$ follow the path $M-\Gamma-X-M$ (Fig. \ref{2D_periodic}). The numerical results were obtained using the two numerical approaches for extracting the smallest eigenvalues, the implicitly restarted Lanczos method and the subspace iteration method, both providing identical results. The number of discretization points used in the numerical approach has to provide sufficient degrees of freedom to the system, accounting for longitudinal and transversal modes in more than one dimension. This number of degrees of freedom has to be at least equal to the number of eigenvalues to be obtained. The diagram in one dimension (Fig. \ref{dis_hom}a)  contains 6 eigenvalues and the discretization uses 6 points. The diagram in two dimensions uses a discretization of 4$\times$4 points and the smallest 12 eigenvalues are represented in  Fig. \ref{dis_hom}b.
In both cases, the resulting frequencies $\omega$ are normalized by $\Omega_0 = L/2\pi c_0$ with $c_0$ the wave speed given by $c_0=\sqrt{E/\rho}$ in one dimension and by the longitudinal speed $c_0=\sqrt{C_{1111}/\rho}$ in 2D.
The results in Figure \ref{dis_hom} show that the proposed numerical technique provides the exact behavior of the homogeneous medium both for one and two dimensional cases. This trivial result confirms the ability of the numerical schemes for eigenvalue extraction to provide the correct ordering of the eigenmodes.

\begin{table}[h!]
\caption{Parameters of the composite material}
\centering
\begin{tabular}{| c|c|c|}
 \hline
& Matrix & Particles \\
 \hline
Composite 1: & Epoxy & Carbon \\
& $E=4.507$GPa, $\nu=.4$  & $E=230$GPa, $\nu=.3$ \\
& $\rho$=1200Kg/m$^3$ & $\rho$=1750Kg/m$^3$ \\
 \hline
  \hline
Composite 2: & Aluminium & Tungsten \\
& $E=74.44$GPa, $\nu=0.334$,  & $E=388.46$GPa, $\nu=.283$ \\
& $\rho$=2692Kg/m$^3$ & $\rho$=19300Kg/m$^3$ \\
 \hline
\end{tabular}
\label{table:1}
\end{table}

\subsection{Multilayer materials}
The second validation case corresponds to obtaining the dispersion diagram in a multi-layer composite for wave propagation perpendicular to the layer surfaces. This problem can be reduced to a 1D heterogeneous medium containing two phases in which an analytical solution is available. The unit cell is formed by a bi-layer of epoxy (phase 1) and carbon (phase 2) whose properties are given in Table \ref{table:1}. The microstructure is defined by a layer of epoxy located in  $0<x<L/2$ and a carbon layer in $L/2<x<L$. 

The analytical solution for this problem is given in \cite{YangLee1974,Sun1968} and the eigenvalues for each given value $k$ correspond to the different solutions $\omega$ of the non-linear equation 
\begin{equation}
\cos(kL) =\cos(\frac{\omega L}{2c_1})\cos(\frac{\omega L}{2c_2})+
               \frac{1+p^2}{2p}\sin(\frac{\omega L}{2c_1})\sin(\frac{\omega L}{2c_2})
          \label{eq:laminate}
\end{equation}
with $p=\sqrt{\frac{E_2 \rho_2}{E_1\rho_1}}$, $c_1=\sqrt{\frac{E_1}{ \rho_1}}$ and $c_2=\sqrt{\frac{E_2}{ \rho_2}}$

\begin{figure}[h!]
\begin{overpic}[width=.49\textwidth]{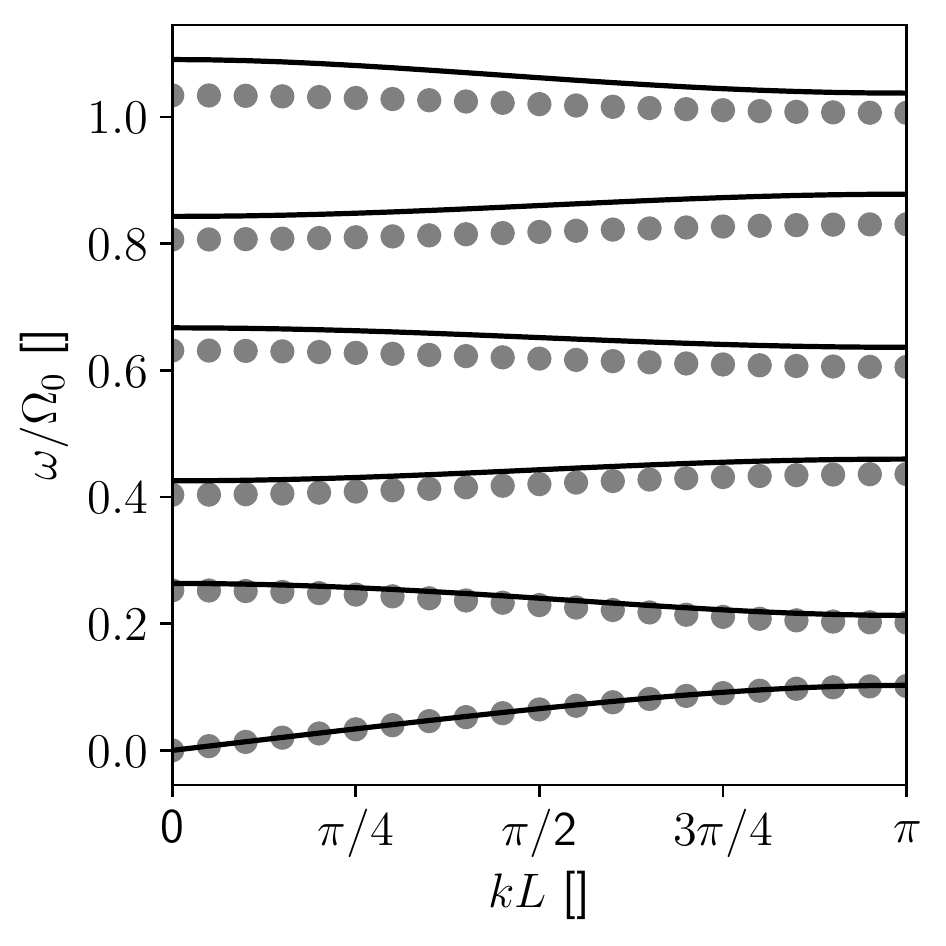}
\put(30,175){$n=16$}
\end{overpic}
\begin{overpic}[width=.49\textwidth]{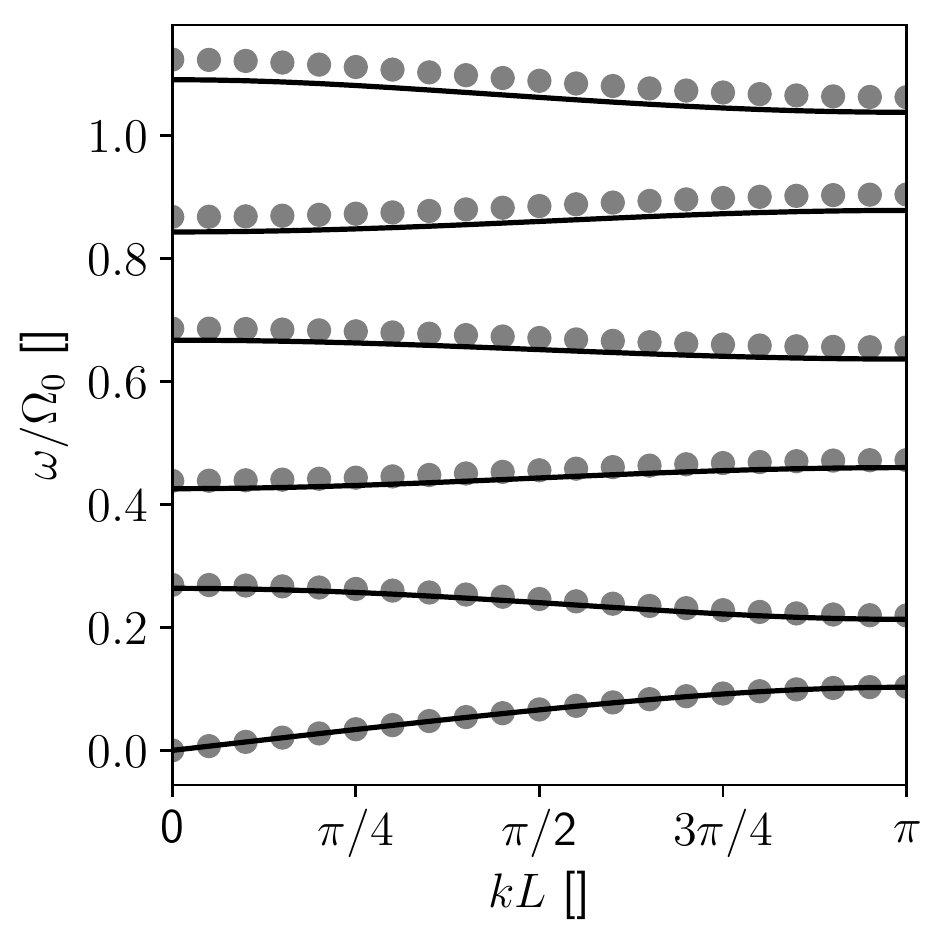}
\put(30,175){$n=32$}
\end{overpic}
\begin{overpic}[width=.49\textwidth]{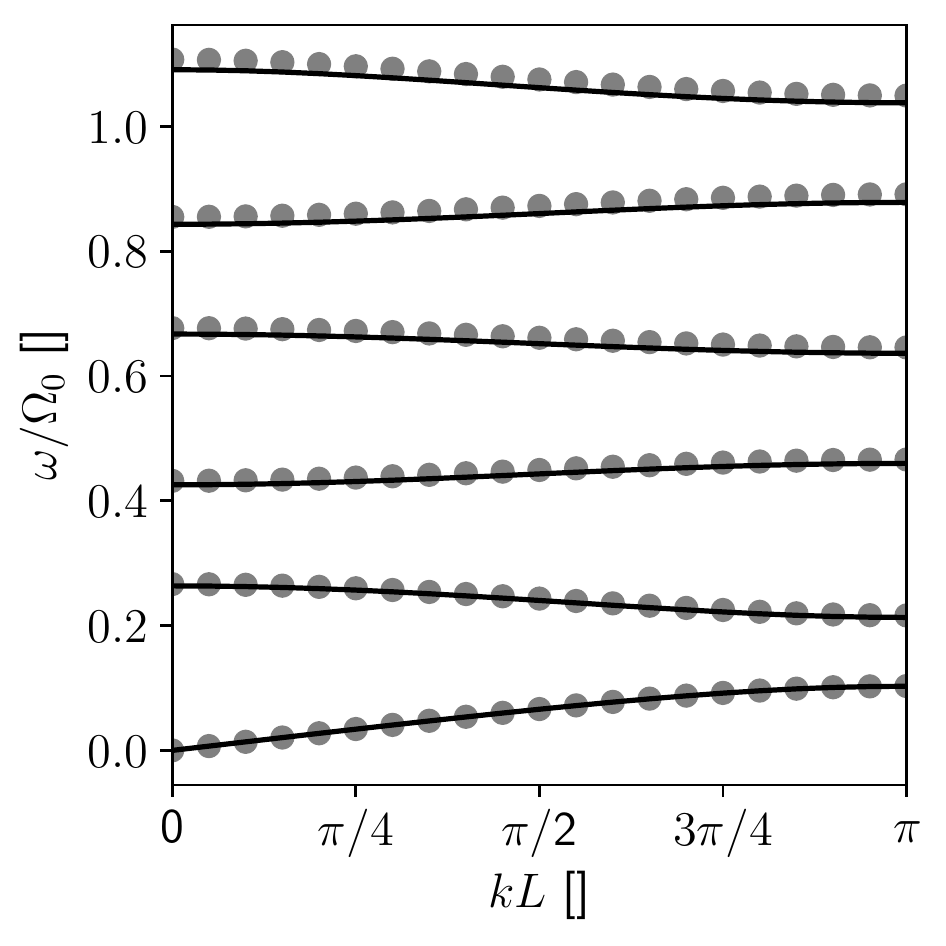}
\put(30,175){$n=64$}
\end{overpic}
\begin{overpic}[width=.49\textwidth]{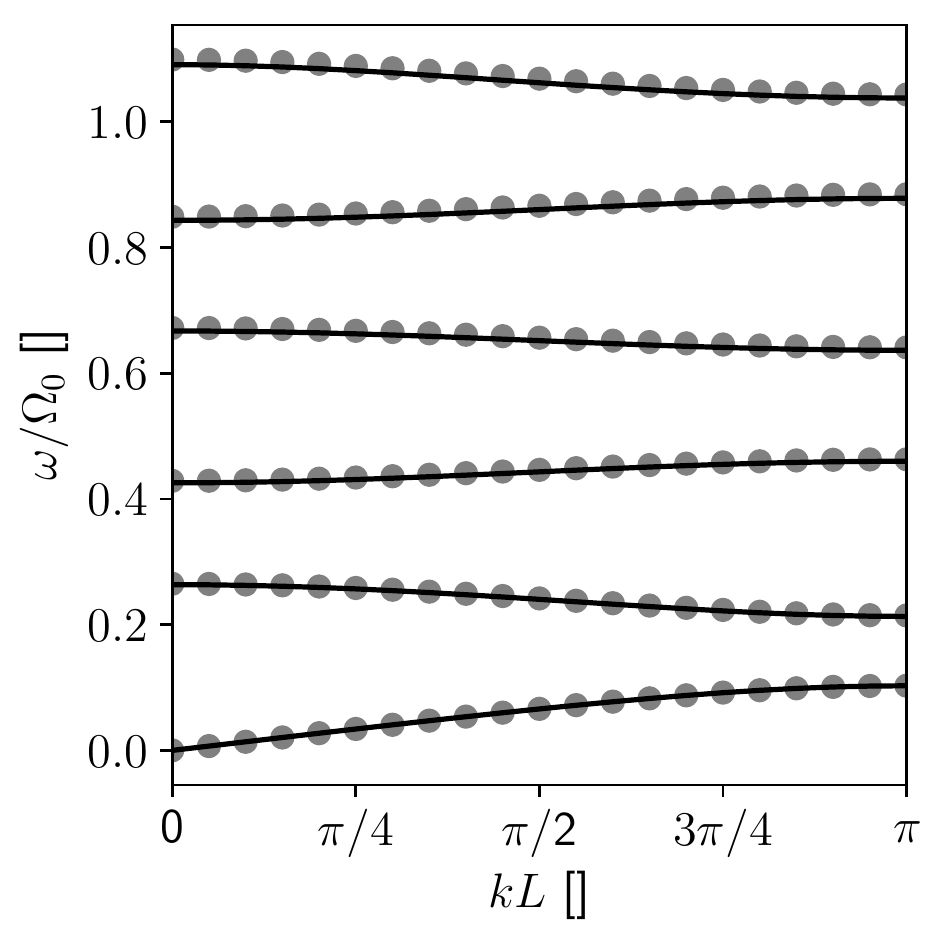}
\put(30,175){$n=256$}
\end{overpic}
\caption{Simulation results of plane wave propagation of a 1D bi-material, Dispersion relation for $n=16,32,64,256$ points}
\label{dis_het}
\end{figure}

\begin{figure}

\begin{center}
\includegraphics[width=.6\textwidth]{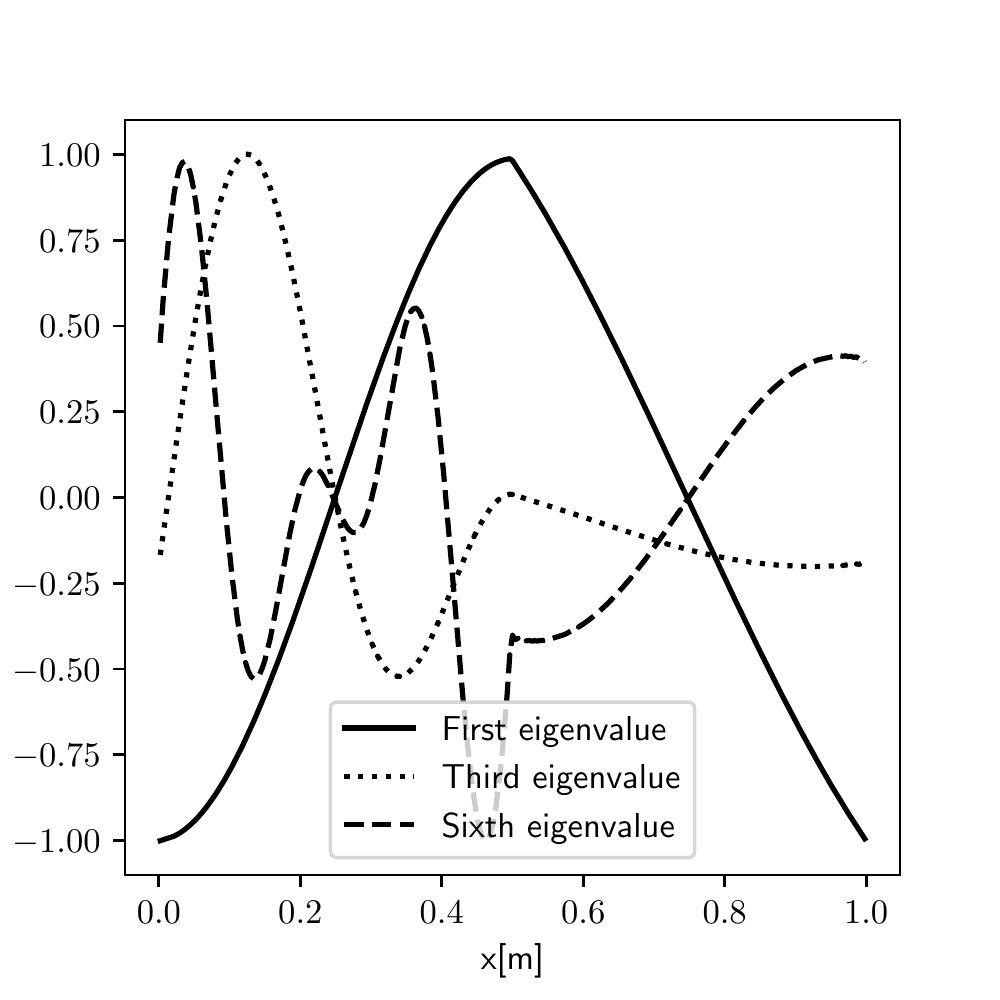}
\end{center}
\caption{Periodic normalized wave shapes $U(x)$ obtained in the unit cell of the 1D bi-layer composite for eigenvalues 1,3 and 6 and  $k=\pi/L$ }\label{eigenvector_1D}
\end{figure}

The problem is solved numerically using the proposed method and four different power-of-two spatial discretizations, from $n=16$ to $n=256$, in order to analyze the effect of the discretization on the resulting diagram. Wave vectors are taken $0\leq k\leq \pi/L$. The restarted Lanczos method was used to solve the eigenvalue equation. The results obtained for the four different discretizations are represented in Fig. \ref{dis_het}, together with the analytical results obtained using eq. (\ref{eq:laminate}). \rev{In the graphs the normalization factor is $\Omega_0 = L/2\pi c_0$, where $c_0$ corresponds to the wave velocity in an equivalent homogeneous medium, $c_0=\sqrt{\overline{E} / \overline{\rho}}$ where $\overline{E}=.5E_1+.5E_2$ and $\overline{\rho}=.5\rho_1+.5\rho_2$ are the homogeneized Youngs modulus and density respectively using the value of the phase properties in Table \ref{table:1}.}
The obtained dispersion relation (Fig. \ref{dis_het}) does not correspond anymore to a linear relation between angular frequency and the wave number (constant slope of the curves), and depends non-linearly on the particular wave number. Moreover, several band gaps are found, indicating a range of frequencies for which corresponding waves will not transport elastic energy in the material. Concerning the comparison with the analytical result, it can be observed that the first eigenvalues are exactly captured for any discretization of the domain, but the higher eigenvalues considered here are only accurately reproduced using a larger number of points in the discretization. In particulxar, to accurately predict the sixth eigenvalue, $n>256$ points are needed. It should be noted that in this problem the microstructure is exactly represented for any value of $n$, so the need of using a large number of points for accurate values in higher frequencies is due to the need of an accurate representation of the complex shape of the corresponding eigenvectors. This result indicates that the classical Fourier series analysis to compute dispersion diagrams \cite{Vasseur_1994} using a small number of modes is inaccurate, even if the microstructure can be represented exactly for simple geometries.  Moreover, the use of a larger number of modes in these approaches would improve the results, but lead to large dense matrices which eigenvalues cannot be extracted efficiently.  The complex shape of the eigenvectors obtained, corresponding to the periodic displacement $U(x)$ in the cell is represented in Fig. (\ref{eigenvector_1D}) for the eigenvalues 1,3 and 6 corresponding to $k=\pi/L$ and using 256 points.
No analytical expressions are available for heterogeneous materials for higher dimensions so, in next sections, solutions obtained with the proposed method will be compared with results obtained using alternative numerical approaches.

\subsection{Long fibre reinforced composite}

A long fibre reinforced composite will be studied using the proposed methodology, and the solutions will be compared with the results obtained in \cite{Vasseur_1994} using the Fourier series approach. The composite is formed by an elastic epoxy matrix reinforced with carbon long fibers, with  the properties of both phases given in Table \ref{table:1}. Two types of fibres are considered: with circular and rectangular sections \cite{Vasseur_1994}. The material with circular fibre section contains a volume fraction $55\%$ of fibres, and the one with square fibres contains $65\%$. The combination of high volume fractions and large contrast in phase properties results in the formation of band-gaps in which the waves do not propagate \cite{Vasseur_1994}. 

\begin{figure}
\begin{overpic}[width=.51\textwidth]{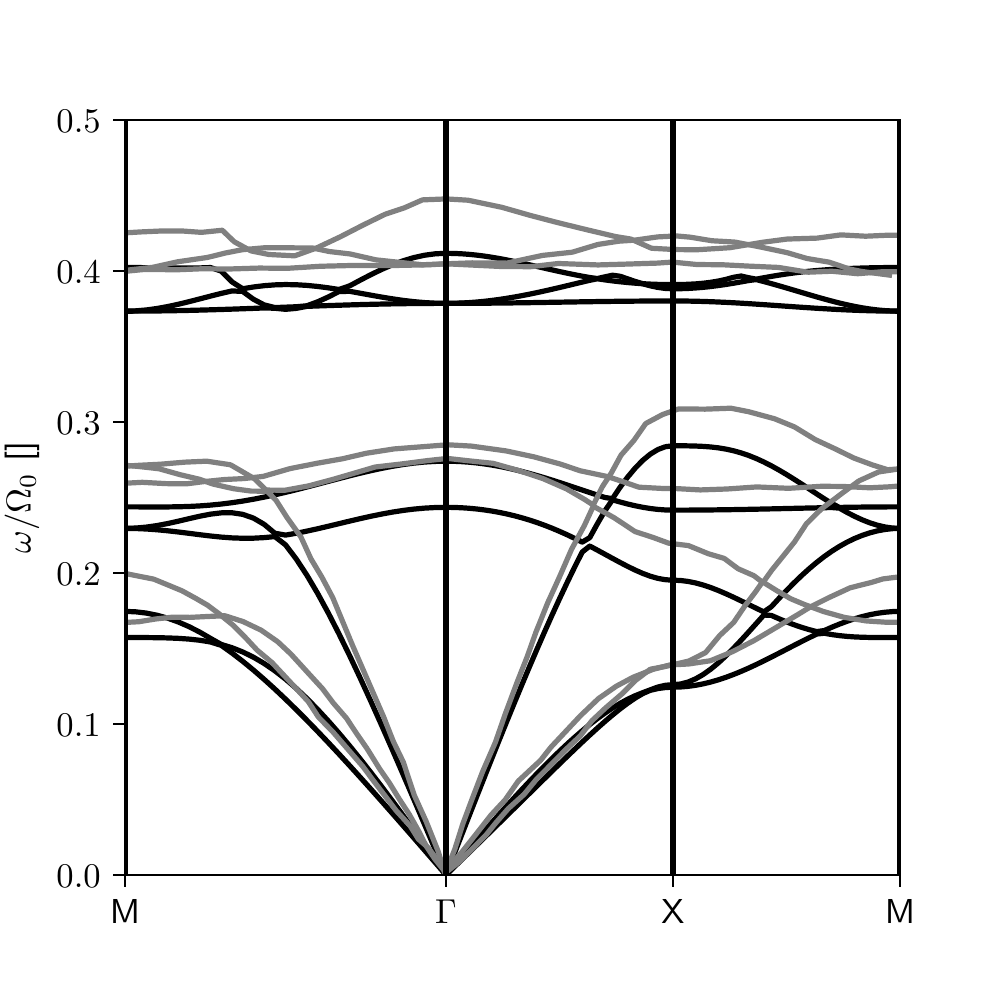}
\put(25,170){(a)}
\end{overpic}
\begin{overpic}[width=.51\textwidth]{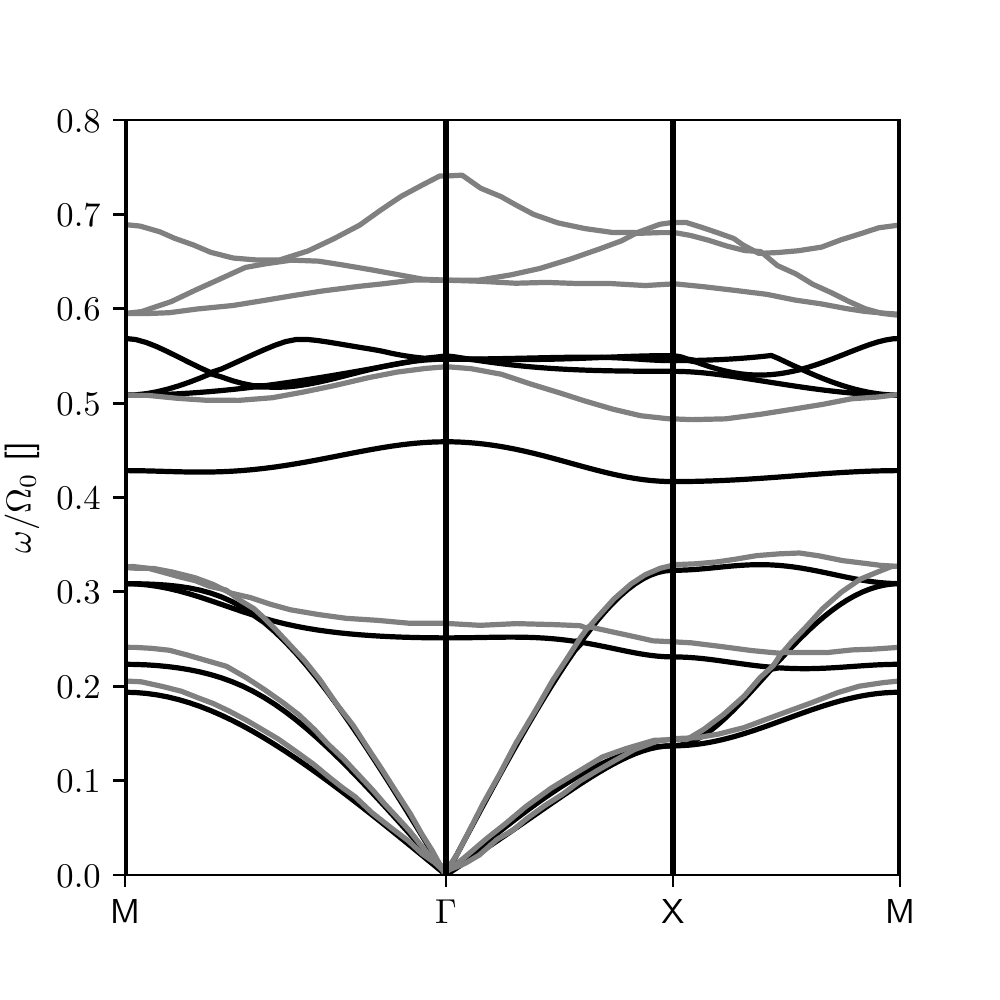}
\put(25,170 ){(b)}
\end{overpic}
\caption{Dispersion diagrams of an epoxy matrix \rev{reinforced} with a square arrangement of long carbon fibers (a) circular section and volume fraction 55$\%$ (b) square section and volume fraction 65$\%$. In black, the present FFT approach, in grey the results obtained in \cite{Vasseur_1994} using Fourier series}
\label{dis_composite}
\end{figure}

The dispersion diagrams  have been simulated for 90 wave vectors $\mathbf{k}$ in the transverse section, $\mathbf{k}=[k_x,k_y,0]$, regularly sampled in the path $M-\Gamma-X-M$ (Fig. \ref{2D_periodic}). A three dimensional model was used, and the unit cells were discretized using a grid with $128\times 128\times 1$ points. The use of a single point in the longitudinal axis (parallel to the fibres) is due to the uniformity of the microstructure in that direction, and was also included in the original simulations in \cite{Vasseur_1994}.  The restarted Lanczos method was chosen for the eigenvalue extraction since it was shown that in this case it was faster than the subspace iteration method. The numerical results including 10 eigenvalues are represented in Figure \ref{dis_composite} together with the results obtained in  \cite{Vasseur_1994}  using a Fourier series approach with 6 modes. \rev{As in the previous cases, the graphs are normalized with a factor $\Omega_0 = L/2\pi c_0$, where $c_0$ here corresponds to the transverse wave velocity in a monolithic medium, $c_0=\sqrt{\overline{G} / \overline{\rho}}$ where $\overline{G}=.5G_1+.5G_2$ and $\overline{\rho}=.5\rho_1+.5\rho_2$ are the shear modulus and density of the monolithic medium respectively, and $G_1,G_2,\rho_1$ and $\rho_2$ are taken from Table \ref{table:1}.}

Figure \ref{dis_composite} shows how the results obtained using the present approach qualitatively reproduce the results obtained in  \cite{Vasseur_1994} exhibiting a band gap between 5th and 6th eigenvalues. However, quantitatively, our results differ from the values obtained in \cite{Vasseur_1994}, with the difference increasing for the largest eigenvalues. As shown in the 1D case, this difference is due to the small number of plane waves used the Fourier series approach in \cite{Vasseur_1994}. It can also be observed that the agreement between both approaches is fairly good for the lowest bands, also in agreement with the 1D case.

A remarkable difference between Fourier analysis and the present method is that in the former approach \cite{Vasseur_1994}, the geometry is accounted through a structure factor that, for simple cases of two-phase materials such as circular and square cells, has an analytical expression. On the contrary, in our approach the geometry of the cell might have any arbitrary shape and contain many different phases since the microstructure is accounted explicitely by a voxelized representation.

\subsection{Effect of the discretization}
\begin{figure}[h!]
\begin{overpic}[width=.49\textwidth]{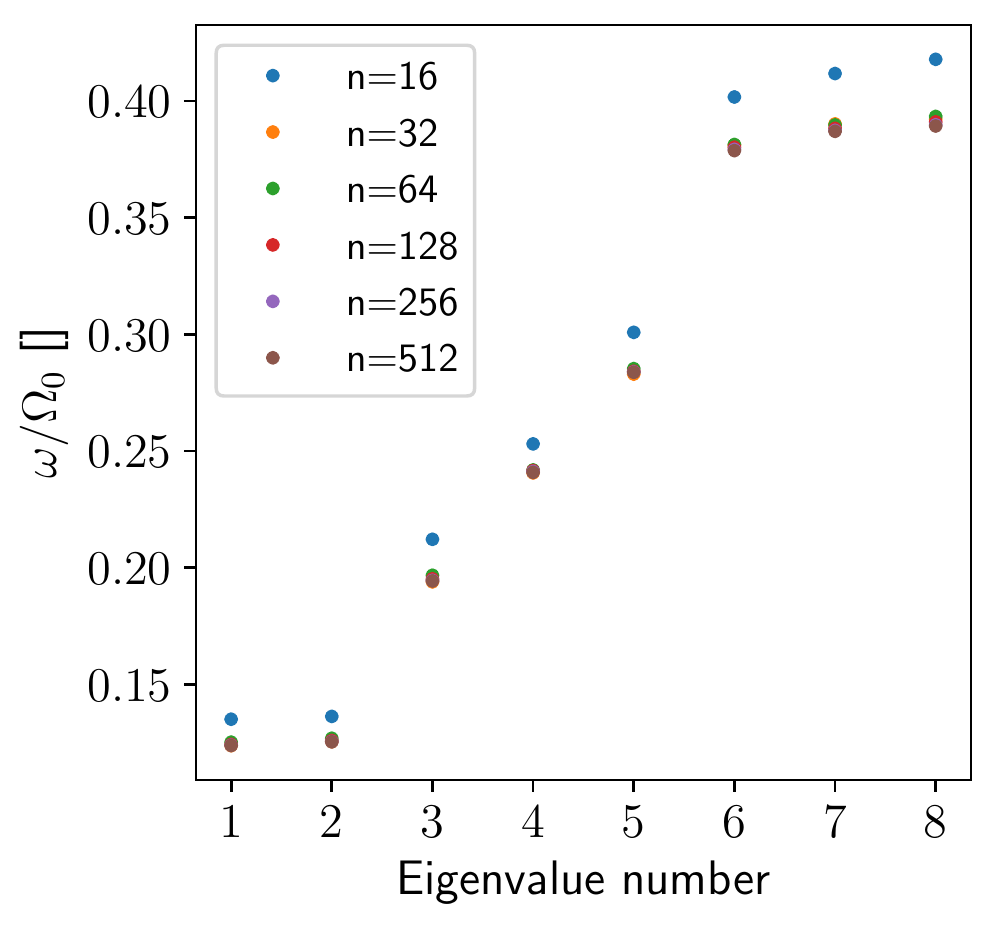}
\put(30,163){\small (a) circular section}
\end{overpic}
\begin{overpic}[width=.49\textwidth]{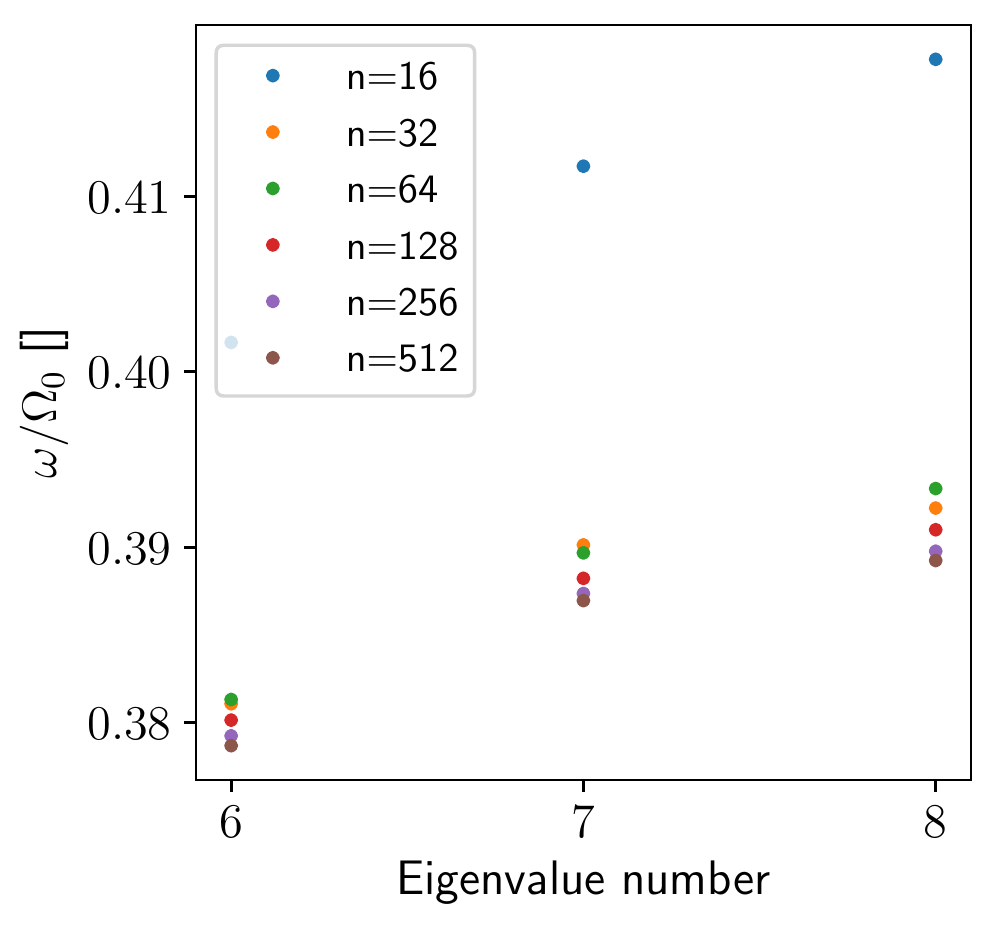}\\
\put(30,163){\small (b) circular section}
\end{overpic}
\begin{overpic}[width=.49\textwidth]{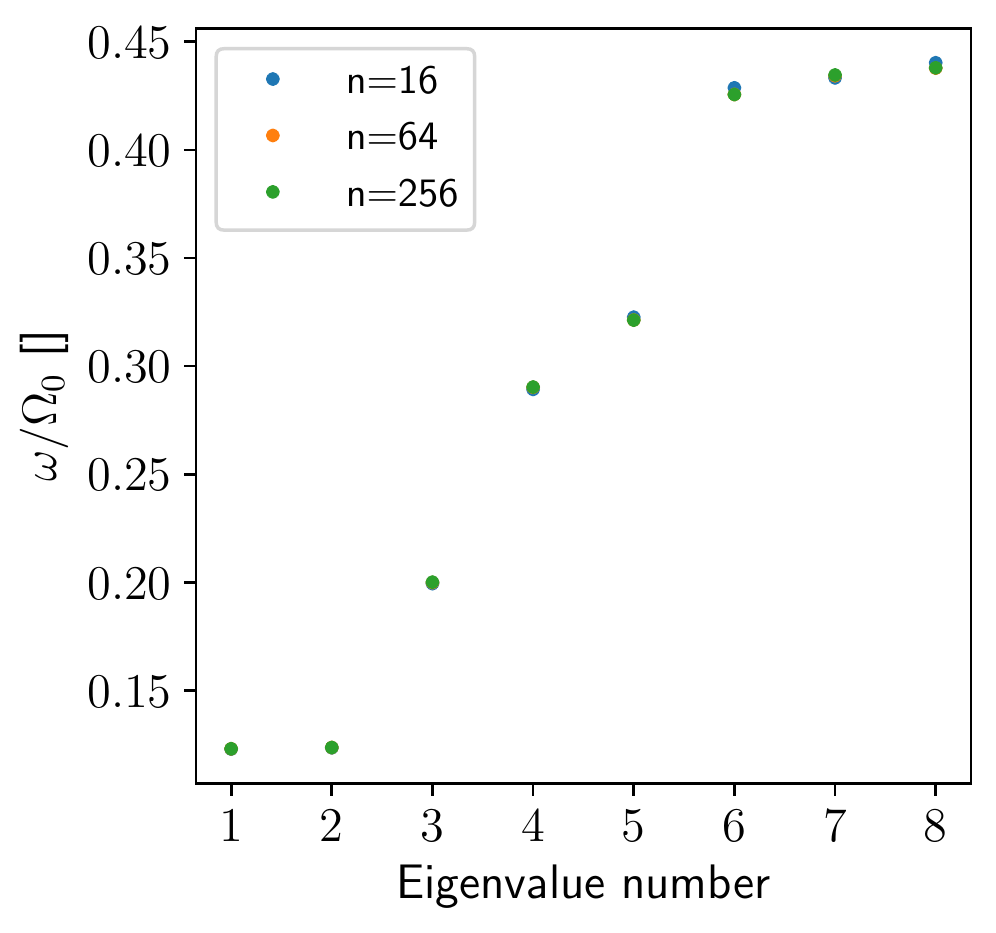}
\put(30,163){\small (c) square section}
\end{overpic}
\begin{overpic}[width=.49\textwidth]{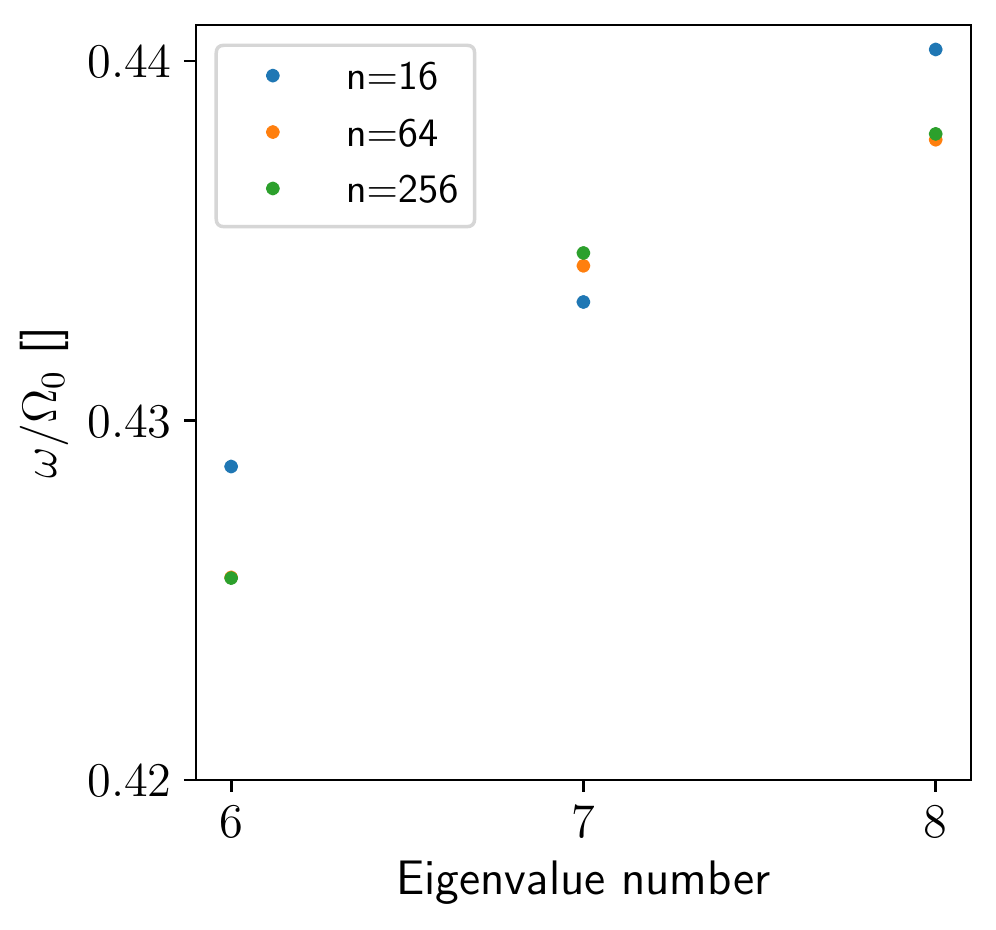}
\put(30,163){\small (d) square section}
\end{overpic}
\caption{Effect of the number of points used for the discretization in the frequency for in  $\mathbf{k}=[ \pi/L,0,0]$ (a,b) the composite with circular fibers (c,d) the composite with square section and volume fraction 56.25$\%$. }
\label{numpoints}
\end{figure}

The long fiber composite problem of the previous section has been solved using different levels of discretization to analyze the convergence of the eigenvalues with the number of points used to solve the problem. For the two microstructures considered, a fixed vector number $\mathbf{k}=[ \pi/L,0,0]$ is used, obtaining the first six eigenvalues for a discretization of the cell from $16\times 16\times 1$ to $512\times 512\times 1$. The results are represented in Fig. \ref{numpoints} \rev{using the same normalization factor $\Omega_0$ as in Figure \ref{dis_composite}.}

First, it can be observed that in the case of circular fibers, a discretization of at least $32\times 32\times 1$ is required to have an error below 5\%, while in the case of the square section $16 \times 16\times 1$ is sufficient. The faster convergence of the eigenvalues in the second case is due to the exact representation of the geometry in the case of the square section. In the case of a circular fibers, the voxelized representation of the geometry leads to small differences in the actual volume fraction for very coarse discretization, which is mainly responsible for the differences. The convergence of the eigenvalues with the discretization becomes slower for higher bands, as it happened in the 1D case. Moreover, the convergence is also very different depending on the vector number chosen. In the present case, when the branch $\Gamma-X$ ($\mathbf{k}\propto [1 ,0,0]$) is explored, it can be observed that our approach provides results very similar to the Fourier series approach with 10 modes \cite{Vasseur_1994} (Fig \ref{dis_composite} with $\mathbf{k}$ in the middle between $\Gamma$ and $X$), indicating that a coarse discretization is sufficient. On the contrary, for values of $\mathbf{k}$ between $X$ and $M$ ($\mathbf{k}$ oriented between [1,0,0] and [1,1,0]) the  convergence rate becomes much slower,  indicating that the actual eigenvectors for that branch have a very complex shape and therefore \rev{require} a finer spatial discretization to be accurately represented.

\section{Application to elastic polycrystals}

In this section, the developed methodology will be applied to study the acoustic response of elastic polycrystals. First, an ideal 1D polycrystal will be analyzed to compute the dependence of the wave group velocity as function of the ratio between incident wave length and average grain size. Then, realistic 3D polycrystals will be simulated to study the effect of crystal anistropy and texture on the dispersion diagrams.

\subsection{One dimensional polycrystals}

\begin{figure}
\centering
\includegraphics[width=.9\textwidth]{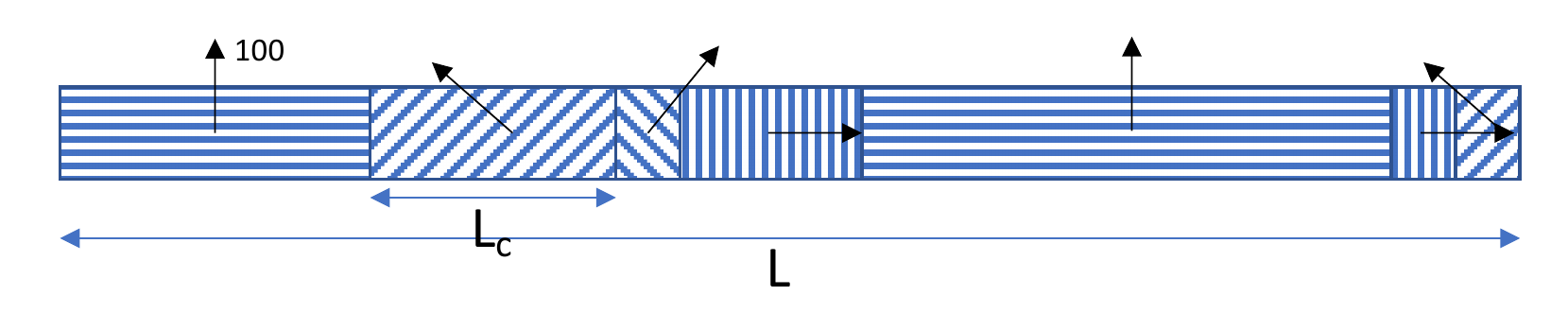}
\caption{Sketch of a 1D random textured polycrystal. Arrows indicate the direction of the [100] direction of each crystal.}\label{fig:1DpolyX}
\end{figure}

An idealized system will be \rev{considered} here to simplify the analysis for incident waves with wave lengths smaller than the periodic domain. The system is a bar composed by $N_c$ crystals where each grain has a different length which is randomly assigned to follow a Gaussian distribution with mean $\bar{D}$. A sketch of the 1D polycrystal is given in Fig. \ref{fig:1DpolyX}. The crystal elastic constants $C_{11}$ and $C_{12}$ as well as the density value are taken from pure Ni, $C_{11}=253$ GPa, $C_{12}=152$ GPa \cite{NeBraCh1952}  and $\rho=8908$ Kg/m$^3$. The constant $C_{44}$ is fixed artificially to set the anisotropy level. For this, the Zener anisotropy ratio $A$, given by
\rev{
\begin{equation}
A=\frac{2C_{44}}{C_{11}-C_{12}},
\end{equation}
is fixed and $C_{44}$ is obtained from previous equation.}

Four levels of crystal anisotropy are considered, A=1 (isotropic single crystal), A=2.45 (which corresponds to the actual Zener ratio in Ni where $C_{44}=124$ GPa) and  A=5 and A=10. The number of crystals is set to $Nc=64$ and the number of voxels to 256. Three grain sizes distributions are considered,  with $\bar{D}=10,50$ and 100$ \mu$m, combined with two standard deviations, 0 (monodisperse case) and $\bar{D}$ (polydisperse case). The resulting lengths of the periodic cells were 0.64 mm, 3.2 mm  and 6.4 mm, respectively. In the 1D model, the only relevant elastic constant is the Young modulus in the direction of the bar, which is obtained for each grain rotating the three dimensional stiffness tensor according to its orientation. A random texture is chosen, so each crystal orientation is randomly generated. 

\begin{figure}
\centering{
\begin{overpic}[width=.32\textwidth]{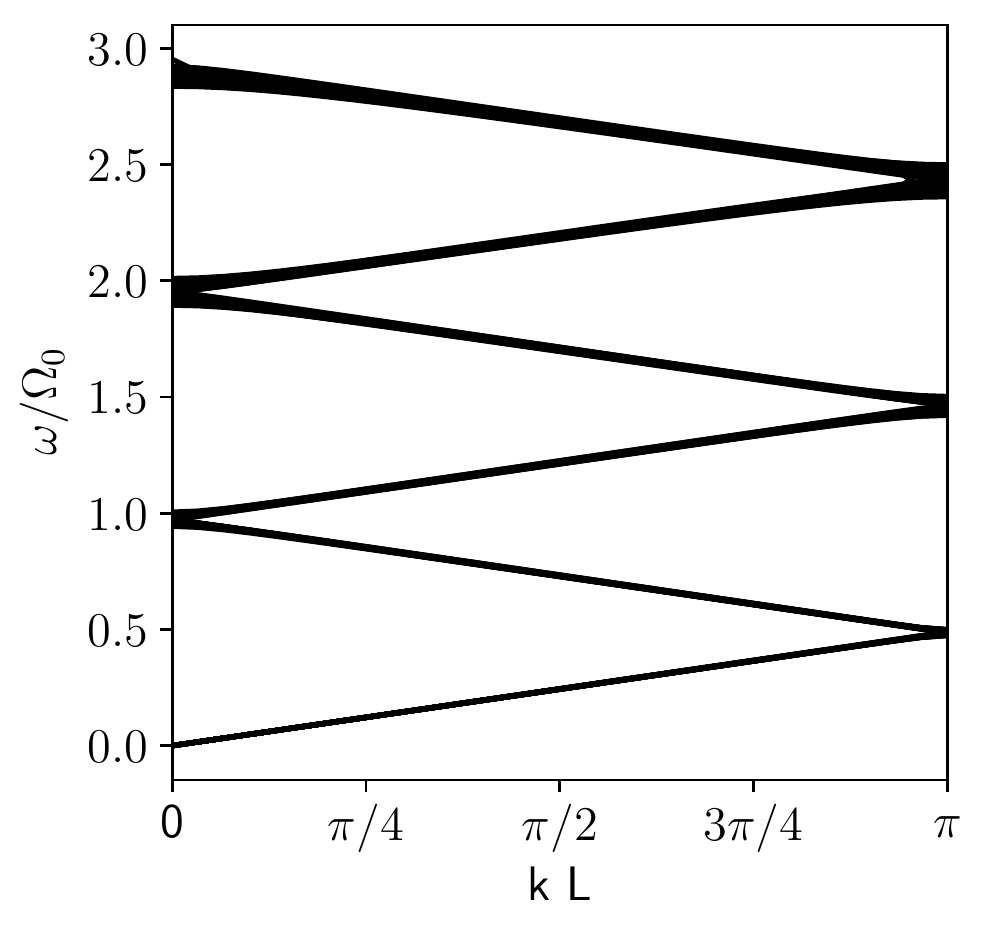}
\put(10,110){\small (a)}
\end{overpic}
\begin{overpic}[width=.32\textwidth]{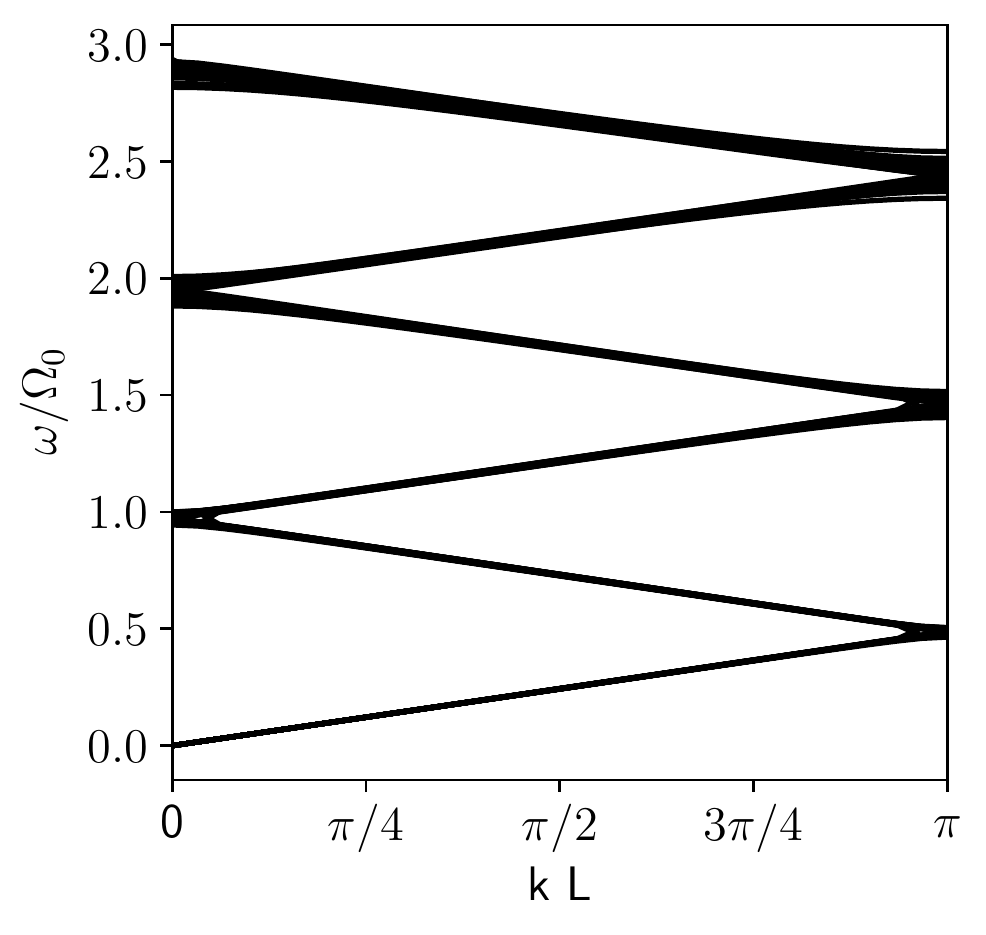}
\put(10,110){\small (b)}
\end{overpic}
\begin{overpic}[width=.32\textwidth]{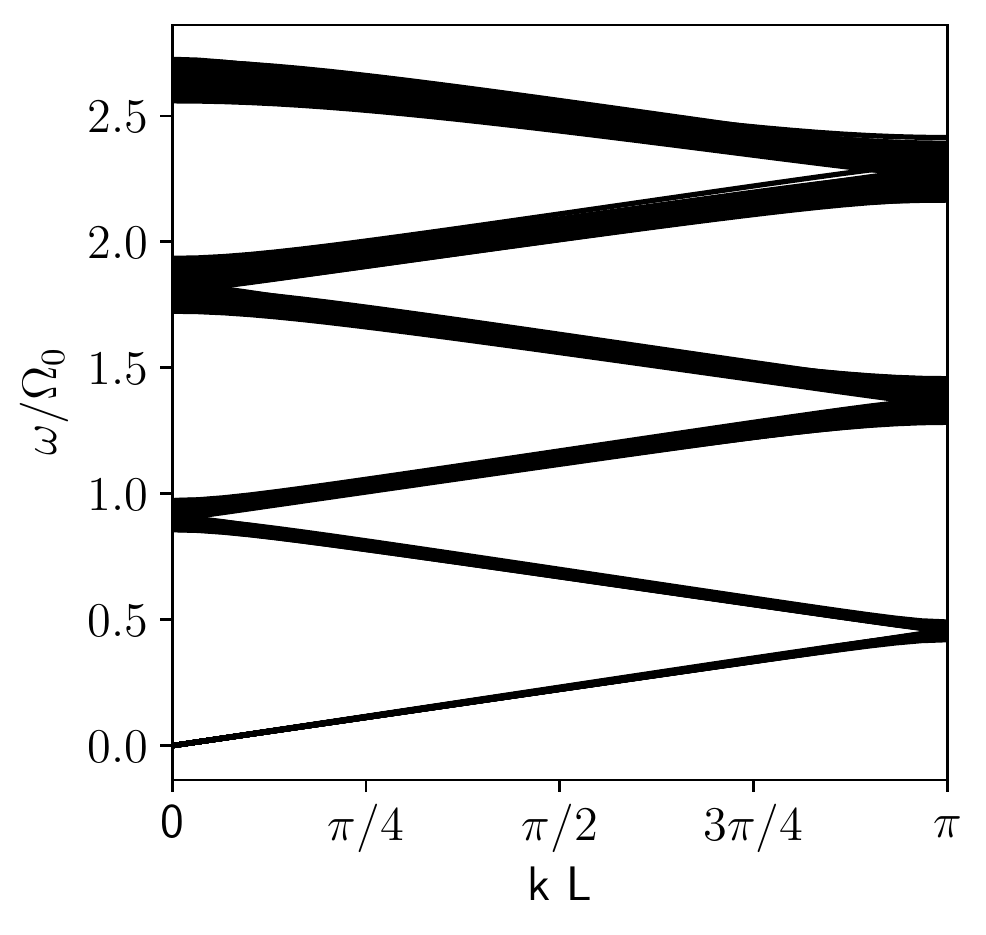}
\put(10,110){\small (c)}
\end{overpic}}
\caption{Dispersion diagram of 40 different 1D polycrystals with $\bar{D}=50$.  (a) Zener parameter =2.45 and monodisperse grain size   (b) Zener parameter =2.45 and polydisperse grain size, (c) Zener parameter =10 and polydisperse grain size}\label{fig:1DpolyX_dispersion}
\end{figure}

The algorithm presented in section 3 is used to compute the dispersion diagram of 40 different random polycrystals with grain size distributions having mean grain size $\bar{D}=50 \mu$m. Three cases are considered, two of them asume real Ni properties with A=2.45 and (a) monodisperse grain size distribution, or (b) Gaussian distribution with standard deviation equal to the average grain size. The third case, (c) corresponds to the Gaussian distribution and an extremely anisotropic single crystals, with A=10. The results are represented in Fig. \ref{fig:1DpolyX_dispersion} where the angular frequency is normalized by \rev{$\Omega_0=L/2\pi c_0$} with $c_0$ defined as the wave speed for an equivalent homogeneous medium with average Young´s modulus \rev{$\overline{E}$.
\begin{equation}
c_0=\sqrt{\frac{\overline{E}}{\rho}}\label{eq:c0}
\end{equation}
}

Fig. \ref{fig:1DpolyX_dispersion} shows that, due to their random microstructures, each polycrystal produced a dispersion diagram slightly different from the rest and, when representing all the diagrams together, the plot appears as an indivudual dispersion diagram, but with the curves thickened. The widths of the curves correspond to the statistical dispersion. The curves corresponding to the first eigenvalue and small $k$ are almost coinciding since they represent incident wave lengths much larger than periodic length and grain size. Differences increase with the eigenvalue number, since for the same $k$ the different eigenvalues correspond to waves with same wave number reduced to the first Brioullin zone $k$, but decreasing wave length. If $p$ is the eigenvalue number, $p=1,2,3...$, then $\omega_p(k)$ is the frequency corresponding to a wave of length $\lambda_p$ given by
\begin{equation}
\lambda_p= \frac{2\pi}{k+\frac{2\pi p}{L}}\label{eq:branches}
\end{equation}
In all cases, the superposition of curves gets thicker near the diagram edges $k=0,k=\pi/L$. This is because the total polycrystal length is equal to an integer number wave lengths for those $k$ values. This effect is an artefact related to the polycrystal's periodicity. It can be observed that differences between polycrystals increase (i.e. curves get thicker) with the grain size distribution range (Fig. \ref{fig:1DpolyX_dispersion}  (a) and (b)) and, for the same grain size distribution, differences increase with crystal anisotropy (Fig. \ref{fig:1DpolyX_dispersion}  (b) and (c)).

\begin{figure}
\begin{overpic}[width=.32\textwidth]{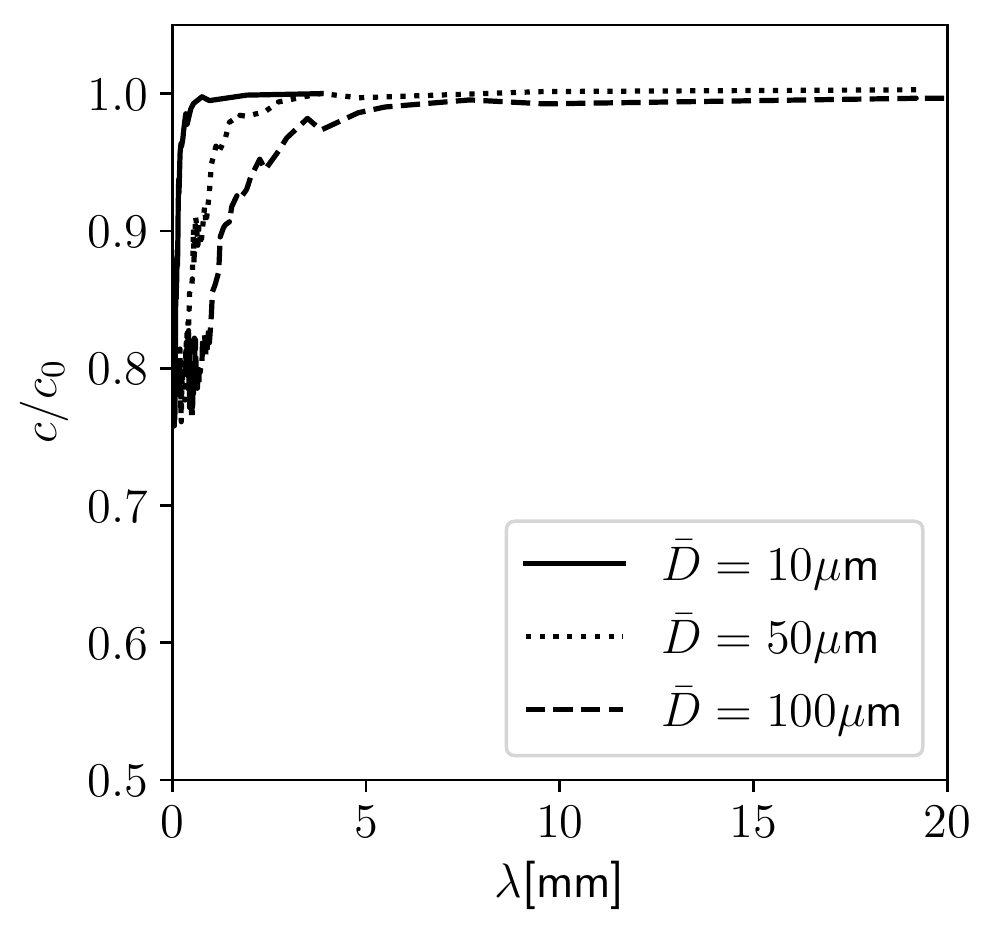}
\put(10,110){\small (a)}
\end{overpic}
\begin{overpic}[width=.32\textwidth]{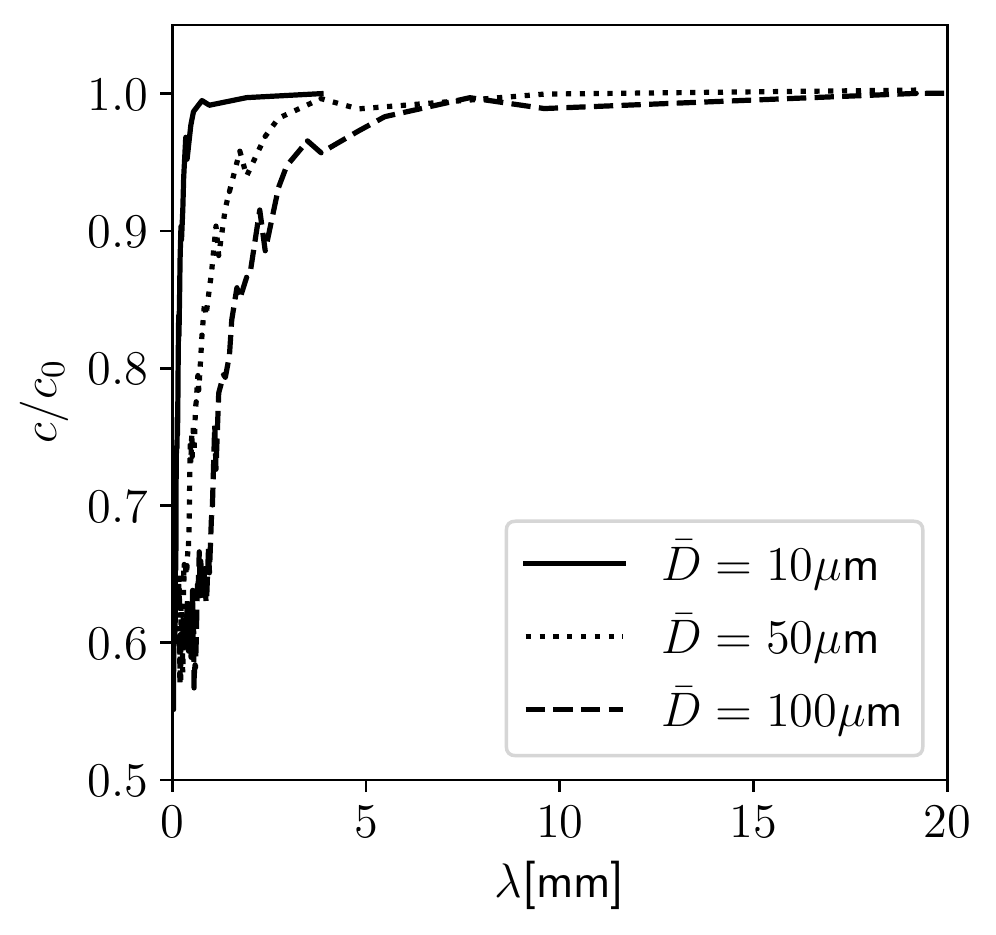}
\put(10,110){\small (b)}
\end{overpic}
\begin{overpic}[width=.32\textwidth]{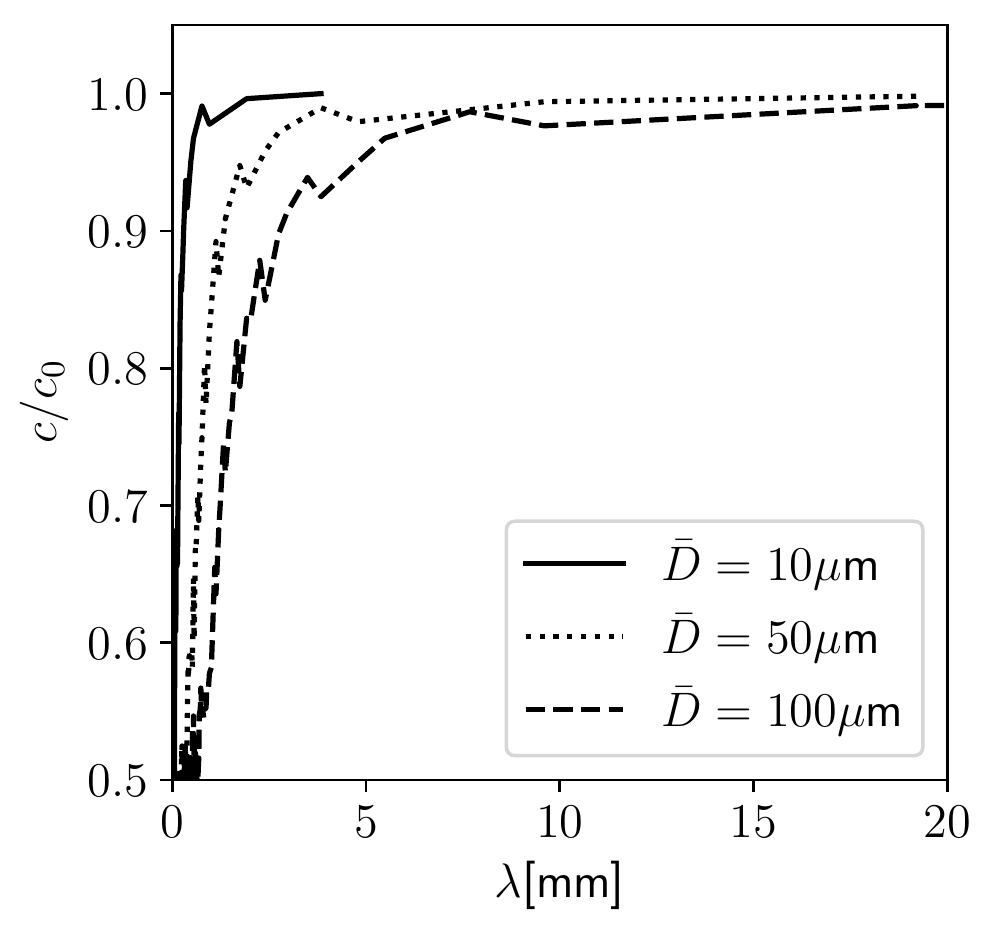}
\put(10,110){\small (b)}
\end{overpic}
\caption{Average normalized wave group velocity in a 1D polycrystal as function of incident wave length for different levels of anisotropy. (a) Zener parameter = 2.45,  (b) Zener parameter = 5, (c) Zener parameter = 10 }\label{fig:1DpolyX_speed_vs_lambda}
\end{figure}

It is well known that the group wave velocity is defined as
\begin{equation}
c_{group}=\frac{\mathrm{d}\omega}{\mathrm{d}k}.\label{eq:c_group}
\end{equation}
In a monolitic material, $c_{group}$ is constant for any wave, but in a heterogeneous medium $c_{group}$ it depends on the incident wave length, if this length is in the order of the heterogeneity size. Therefore, dispersion diagrams can be used to estimate the wave velocity in dispersive media by computing the derivative in eq. (\ref{eq:c_group}) for a particular incident wave. Moreover, in a 1D diagram for a given reduced $k$, is very easy to identify the wavelength with the eigenvalue branch using eq. (\ref{eq:branches}) so computing that derivative in higher branches would provide the wave group velocity for decreasing wave lengths, smaller than the polycrystal periodicity $L$. For a sufficient high eigenvalue number, the wave length would be in the order of the grain size, and for that range it is expected that $c_{group}$ decreases due to interactions with grains. This calculation has been performed for the same 40 polycrystals and the group velocity (eq.\ref{eq:c_group}) has been obtained for $k=\pi /3L$ and $k=2\pi/3L$ and increasing eigenvalues numbers. The average group velocity, normalized by  $c_0$ (eq. \ref{eq:c0}),  has been plotted against the incident wave lengths given by (eq. \ref{eq:c_group}) in Fig. \ref{fig:1DpolyX_speed_vs_lambda} for the polydisperse grain size distribution, and for $A=2.45,5$ and 10.  The case $A=1$ is trivial since it corresponds to a monolithic material and provides a constant $c_{group}$ for every wave, and has not been plotted.
Fig. \ref{fig:1DpolyX_speed_vs_lambda} shows that $c_{group}$ is almost constant for all microstructures, and equal to the wave velocity in a monolithic equivalent media for large wave lengths 

$$\lim_{\lambda \rightarrow \infty} \frac{\mathrm{d}\omega}{\mathrm{d}k} =\lim_{k \rightarrow 0} \frac{\mathrm{d}\omega}{\mathrm{d}k} = c_0 $$

On the contrary, when incident the wave length decreases 
the wave speed drops, showing the dispersive behavior of the polycrystal due to elastic anisotropy. The wave length at which the drop starts depends on the average grain size. This is a very important result since Fig. \ref{fig:1DpolyX_speed_vs_lambda} is a theoretical prediction of the wave attenuation as function of grain size distribution, which, if extended to 3D and adequately calibrated, could potentially allow estimating grain size from acoustic measurements.

\subsection{Three dimensional polycrystals}
\begin{figure}[h]
\centering
\includegraphics[width=.8\textwidth]{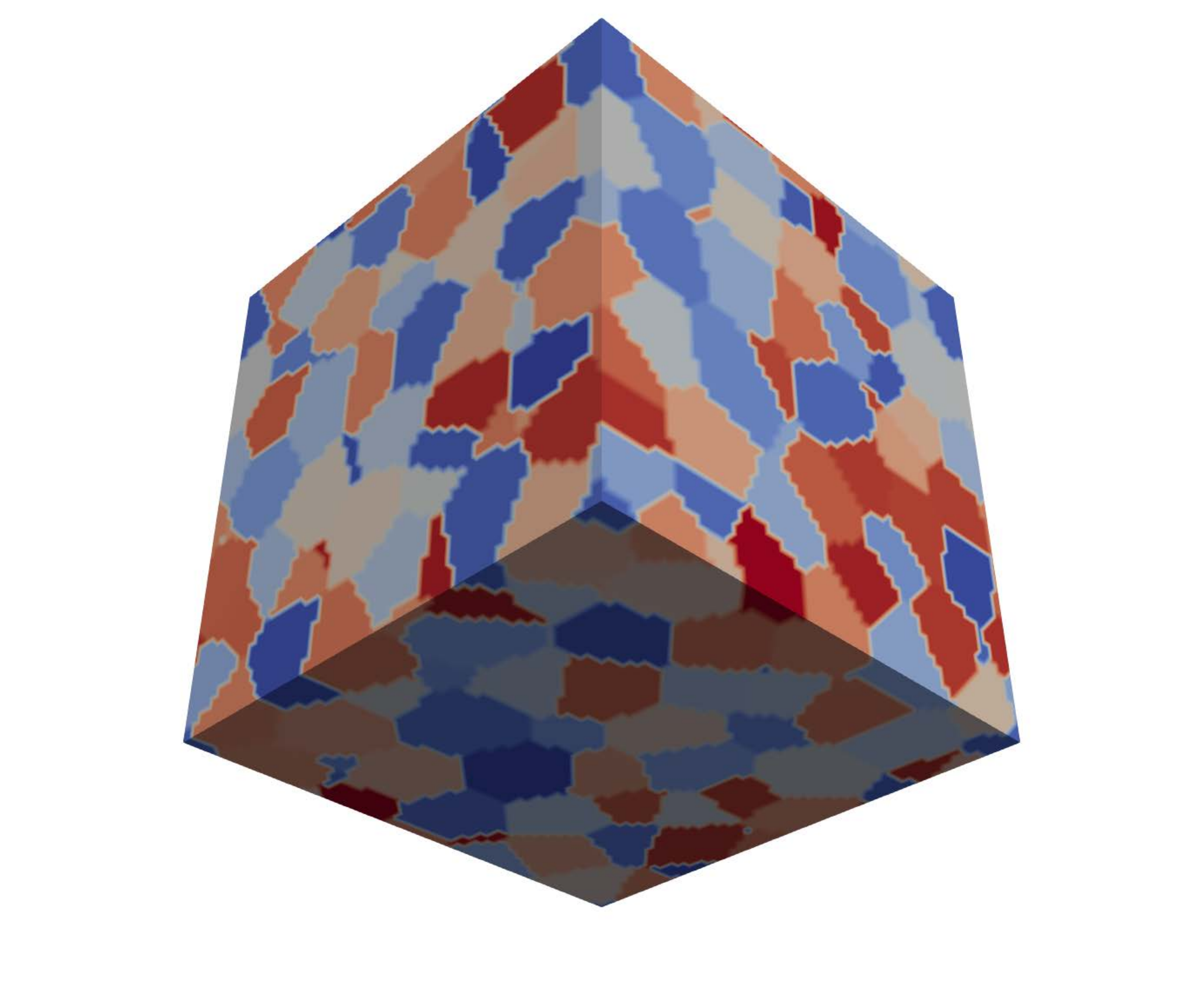}
\caption{Polycrystalline RVE used for the simulations, discretization corresponds to a $64^3$ grid. The colors represent the crystal number.}
\label{fig:polyX}
\end{figure}

In this subsection, the dispersion diagrams for 3-D polycrystalline unit cells are computed using the method described above. Three different periodic and random polycrystalline cells are used, each one discretized in a $64^3$ grid. An example of a polycrystaline RVE is represented in Fig. \ref{fig:polyX}

Each polycrystalline cell contains a set of grains with a  mono-modal grain size distribution. The microstructure is generated using an weighted 
Voronoi tesellation using the FFTMAD code \cite{Lucarini2019a}. The grain diameters follow a log-normal distribution with average diameter of $200 \mu$m, and scatter of $10 \mu$m. The cell size is 1 mm$^3$ and the average number of grains per cell is around 240. The elastic properties correspond to pure Ni with same constants used in the previous section \cite{NeBraCh1952} and a random orientation distribution, texture represented in Fig. \ref{fig:texture}(a). 
FFT-based homogenization has been used to \rev{compute} the effective elastic constants of the polycrystal. The resulting polycrystal's elastic behavior is almost isotropic, with small differences along equivalent directions due to the discrete number of grains considered. The effective Young's modulus was $E=161.4 $GPa, the Poisson's ratio $\nu=0.355$, and the shear modulus $\mu=61.05$ GPa, which correspond to a nearly isotropic material (Zener ratio $A=1.025$). 

\begin{figure}[h]
\centering
\includegraphics[width=.8\textwidth]{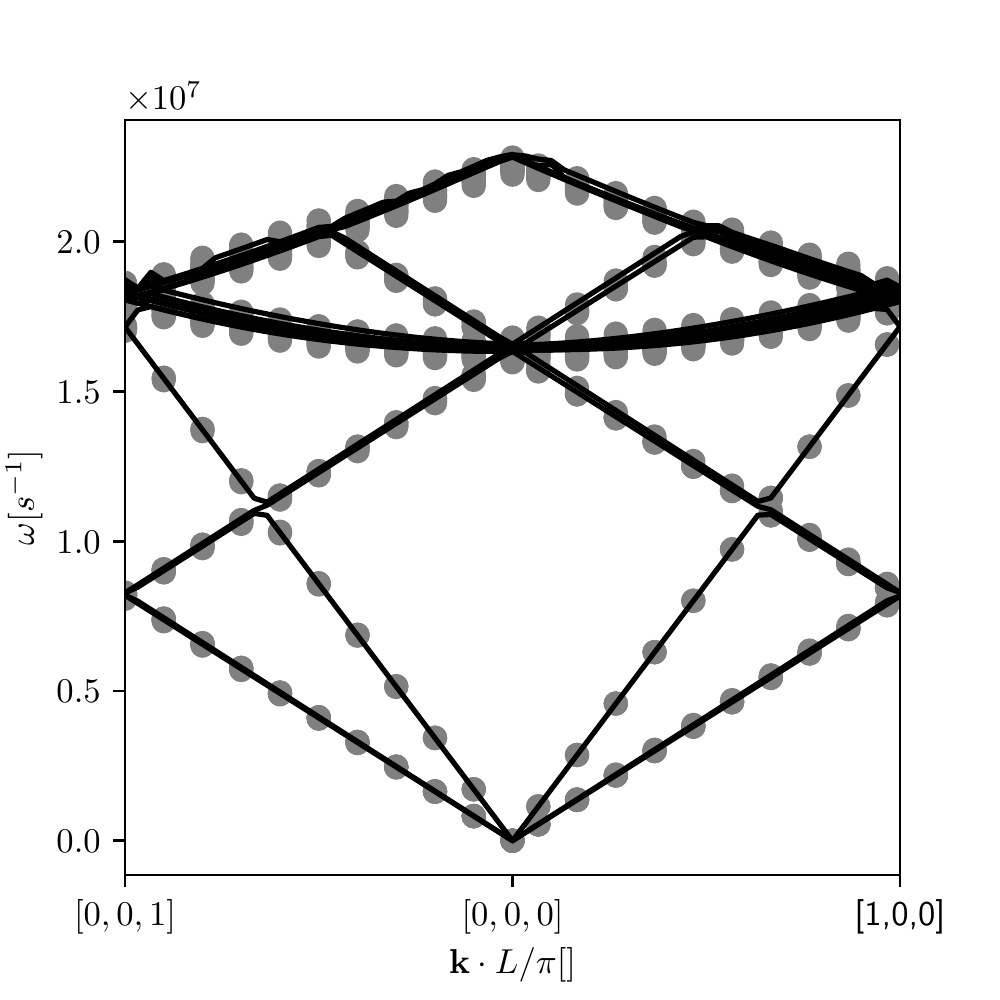}
\caption{Dispersion diagram of 3 random polycrystals (grey dots) and a monolithic metal with equivalent elastic response (black lines).}
\label{fig:polyX_3D_random_64_0dispersion}
\end{figure}

The resulting dispersion diagrams for the three random isotropic polycrystals are represented in Fig. \ref{fig:polyX_3D_random_64_0dispersion}, together with the result for a monolithic material with the homogeneized elastic properties of one of the random realizations. Two wave directions have been explored, oriented in the x-direction ($\mathbf{k} = k[1,0,0]$) and the z-direction ($\mathbf{k} = k[0,0,1]$) with $k\in (0,\pi/L)$. For each direction, 30 wave numbers regularly distributed have been used.
In Fig. \ref{fig:polyX_3D_random_64_0dispersion}, the black lines correspond to the monolithic material with homogeneized properties while grey points represent the response of the three polycrystalline unit cell realizations. The diagram is almost symmetric with respect to $k=[0,0,0]$ and this is due to the random texture that result in nearly isotropic materials, in which all directions are equivalent.  The diagram shows lines corresponding to longitudinal and transverse waves (straight lines) and coupled modes (curved lines). Some of the lines are almost coinciding and cannot be distinguished, since they represent equivalent modes. This happens, for example, for the two perpendicular transverse modes with different eigenvectors but same frequency. Nevertheless, a closer look reveals that the lines are not exactly the same due to the non-strict isotropic response that breaks the eigenvalue degeneracy.

Regarding the difference between the diagram of an homogeneous solid and the results when the polycrystalline character of the microstructures is considered, it is observed that such difference is minimal for low frequencies, for which black lines and grey points (results of FFT-based simulations for the three random microstructures) are almost coinciding. On the contrary, for higher frequencies ($\omega>$15 GHz), the results for the three different numerical realizations show some scatter and do not lie exactly on the line representing the monolithic material. These differences are due to the interaction of the waves with the microstructure, which at low frequencies is negligible, since the waves have wavelengths much larger than the grain size.  These differences will be more clear when analyzing the deformation modes (Fig. \ref{fig:eigenvector_polyX}).

A quantitative analysis that can be carried out with the proposed method consists in obtaining the group wave velocities for low frequencies, i.e.  waves with length much larger than the characteristic size of the microstructure, the grain size in this case. The elastic properties of the homogeneized material will give directly the wave velocity of very large waves (wave number $k\rightarrow 0$), both the transverse and longitudinal velocities, $c_{T}$ and $c_L$, for a wave traveling in a particular direction. For a homogeneous material, $c_L$ is independent on the direction, and transverse wave velocities are unique. For general anisotropy, if the wave vector (propagation direction) is set as $\mathbf{k} = [1,0,0]$, then the longitudinal and transverse wave velocities are given by
\begin{eqnarray}
\label{eq:wave_speed}
c_L=\sqrt{\frac{\bar{C}_{1111}}{\bar{\rho}}} ; \ c_{T1} =\sqrt{ \frac{\bar{C}_{1212}}{\bar{\rho}}}; \
c_{T2} =\sqrt{ \frac{\bar{C}_{1313}}{\bar{\rho}}}\\ \nonumber
\end{eqnarray}
and these velocities can be defined for each direction of $\mathbf{k}$. In the present case, using the values of $\mathbb{C}$ obtained by FFT-based homogenization, the group velocities correspond to longitudinal velocity $c_L=5461.74$m/s and transverse velocities $c_{T1}=2612.62$m/s and  $c_{T2}=2525.52$m/s. The dispersion diagrams can be used to compute the wave velocities for polycrystals as function of the incident wave number, since the wave group velocity for a given frequency corresponds to the derivative of the dispersion relation. In particular, for waves traveling in direction 3,  $\mathbf{k}=[0,1,0]K$ with $K\rightarrow0$ (very large wave length), the transverse wave velocities corresponds to the two smallest eigenvalues  $\omega^1,\omega^2$ and the longitudinal velocity with the third eigenvalue  $\omega^3$
\begin{eqnarray}
\label{eq:wave_speed2}
c_T =\lim_{K\rightarrow 0} \frac{\mathrm{d} \omega^1}{\mathrm{d} K}\\ \nonumber
c_L =\lim_{K\rightarrow 0} \frac{\mathrm{d} \omega^3 }{\mathrm{d} K}
\end{eqnarray}
These derivates are evaluated near the origin using the numerical results shown in Fig. \ref{fig:polyX_3D_random_64_0dispersion}
and the wave velocities obtained are $c_L=5461.76$m/s,  $c_{T1}= 2599.41$m/s and $c_{T2}=2637.71$m/s. As it can be observed, the relative differences between the velocities computed by the two different approaches are below $0.5 \%$ confirming that the influence of the microstructure is negligible for low frequencies.

\subsubsection{Eigenvectors}
The displacement eigenvector corresponding to each eigenvalue $p=1,...,n$ for a given $\mathbf{k}$ is computed in Fourier space, $\hat{\mathbf{U}}^p_{\mathbf{k}}$. The corresponding periodic displacement in real space is a complex-valued field, in which all the points have the same phase $\phi = \arctan (Im (U)/Re (U) )$. The actual displacement field is then obtained 
transforming the eigenvector back to real space as

\begin{equation}
\mathbf{U}^p_{\mathbf{k}}\mathbf{(x)} = e^{-i\phi} \mathcal{F}^{-1} (\hat{\mathbf{U}}^p_{\mathbf{k}}).
\label{eq:rotate_eigenvector}
\end{equation}

\begin{figure}[h!]
\begin{center}
\begin{overpic}[width=.48\textwidth]{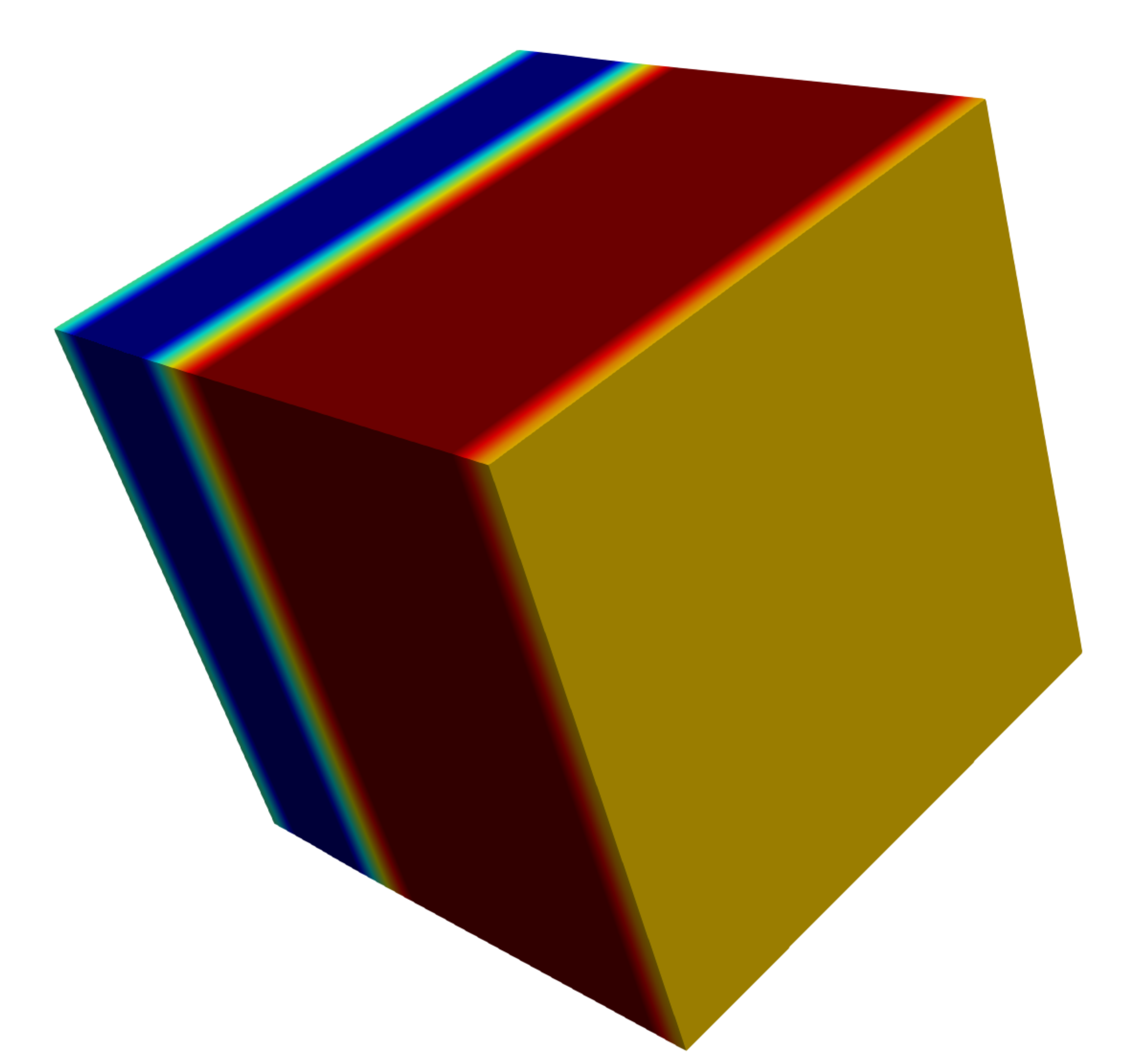}
\put(10,150){(a)}
\end{overpic}
\begin{overpic}[width=.48\textwidth]{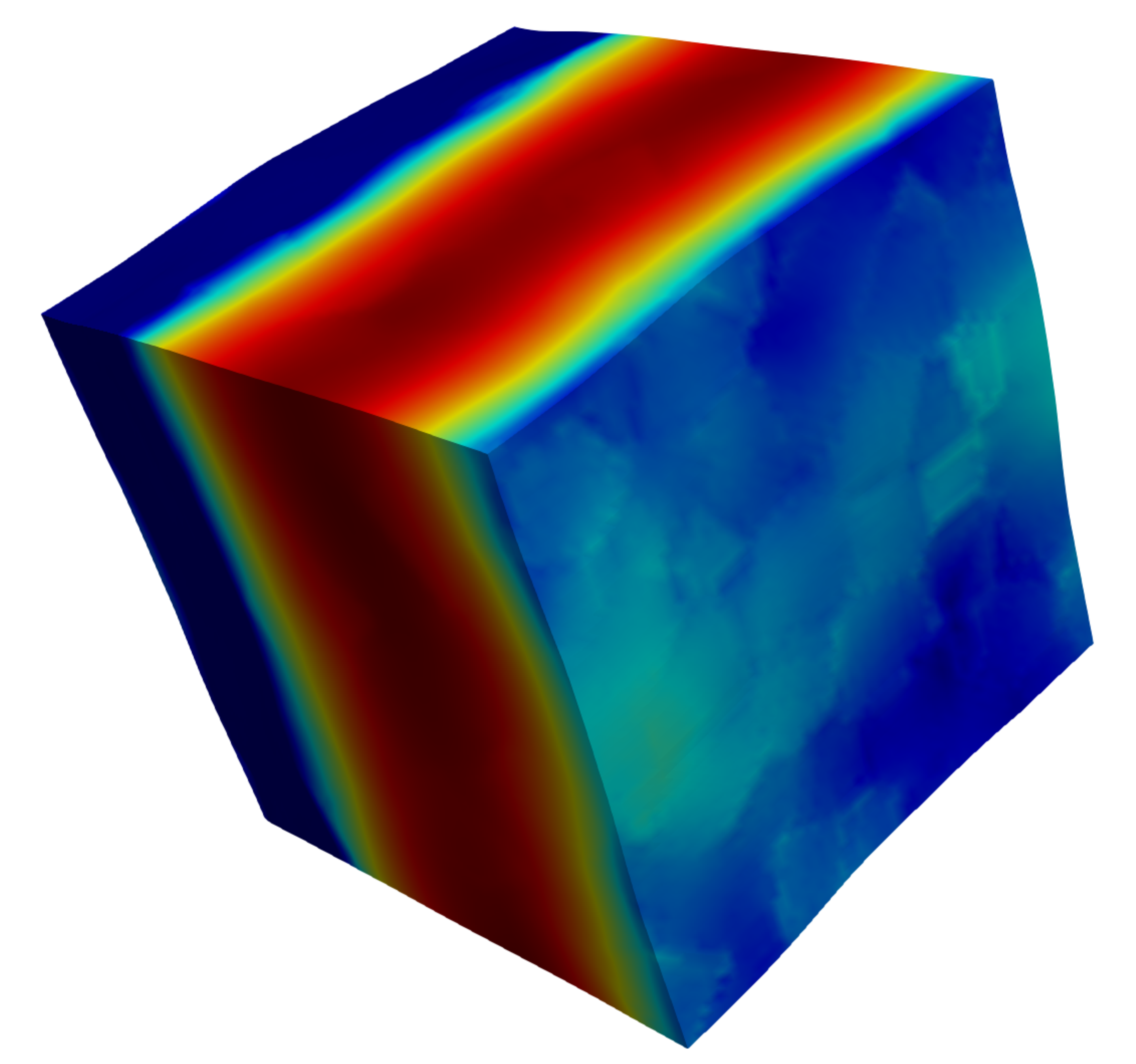}
\put(10,150){(b)}
\end{overpic}
\begin{overpic}[width=.48\textwidth]{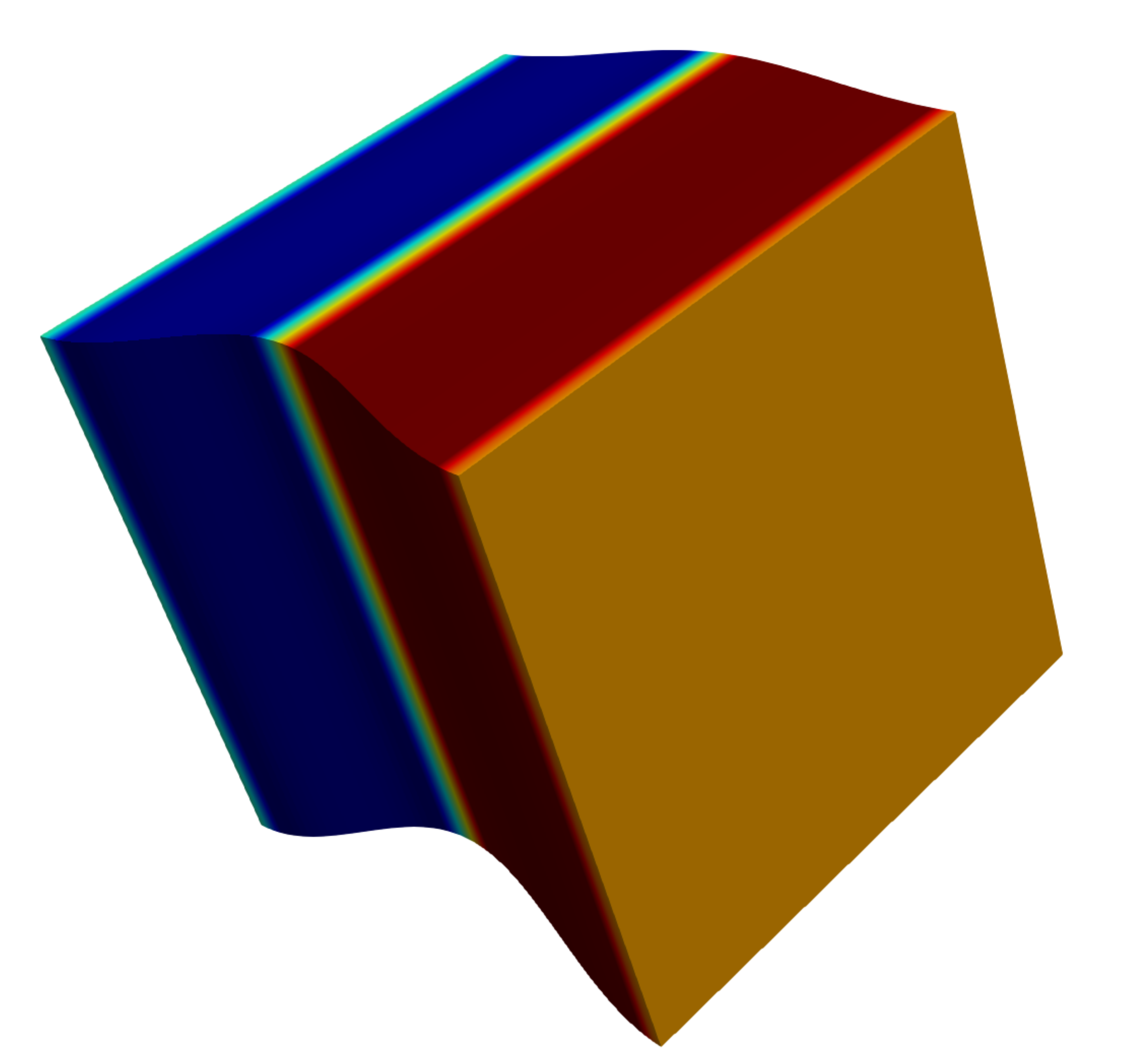}
\put(10,150){(c)}
\end{overpic}
\begin{overpic}[width=.48\textwidth]{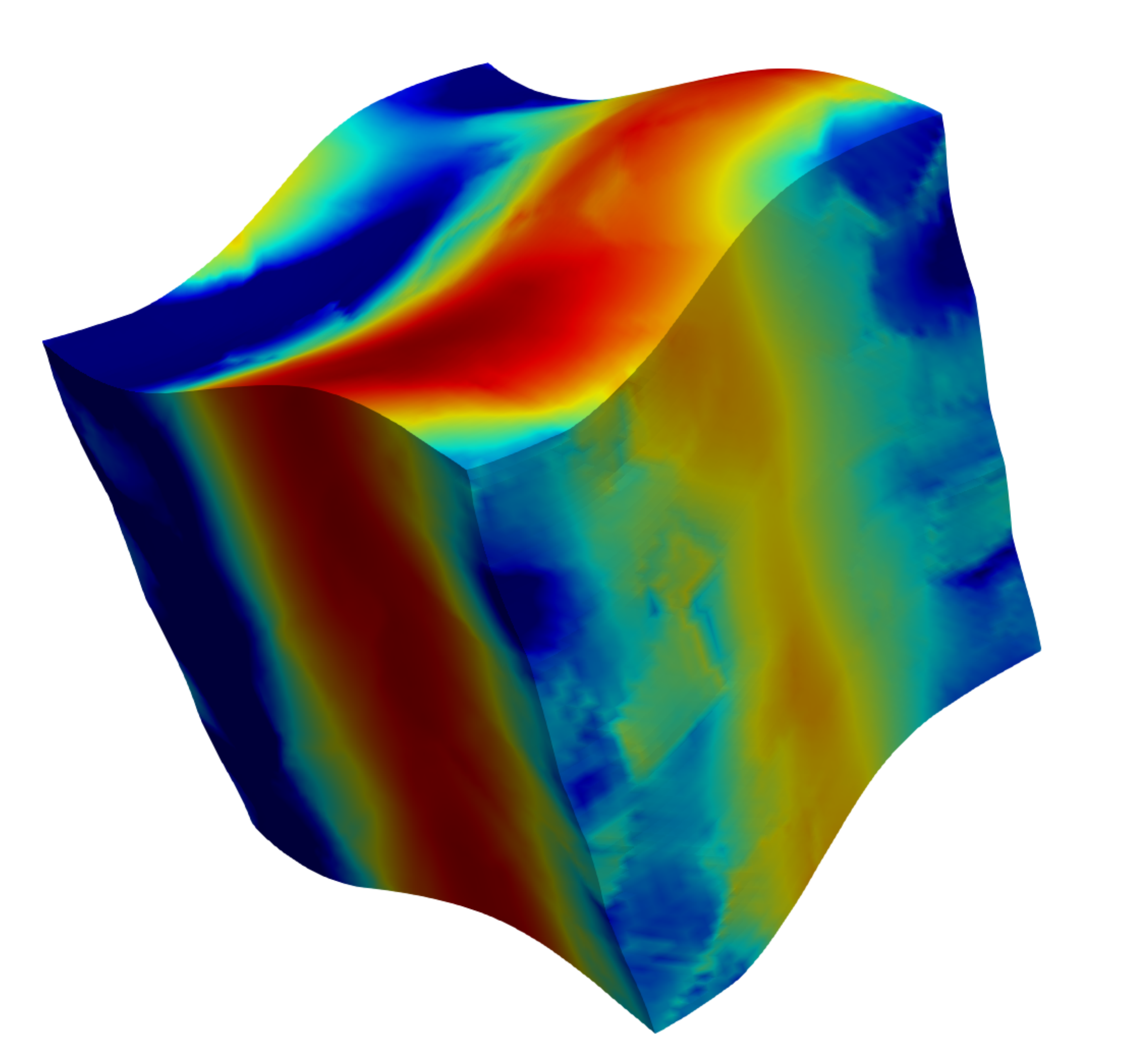}
\put(10,150){(d)}
\end{overpic}
\caption{Bloch eigenvalues of the polycrystal.  (a) $\mathbf{k}=[0.999,0,0] \pi/L$, equivalent homogeneous material. (b) $\mathbf{k}=[0.999,0,0] \pi/L$, random polycrystal.  (c) $\mathbf{k}=[0.999,0,0.999]\pi/L$, equivalent homogeneous material. (d) $\mathbf{k}=[0.999,0,0.999]\pi/L$, random polycrystal. The deformation corresponds to the eigenvalue $\mathbf{U(x)}$ and the colour levels correspond to the strain $\epsilon_{11}=\nabla \mathbf{U(x)}$}\label{fig:eigenvector_polyX}
\end{center}
\end{figure}

The Bloch periodic displacement fields $\mathbf{U}^p_{\mathbf{k}}\mathbf{(x)}$ corresponding to the eigenvalues represented in Fig. \ref{fig:polyX_3D_random_64_0dispersion} have been computed using eq. (\ref{eq:rotate_eigenvector}). The deformed polycrystalline domain corresponding to the eigenvalues $p=6$ for $k=0.999\pi/L[1,0,0]$ and $k=0.999\pi/L[1,0,1]$  are represented in Fig. \ref{fig:eigenvector_polyX}, superimposing as contour plot the value of $\epsilon_{11}=\frac{\partial {U_1}^p_{\mathbf{k}}}{\partial{x_1}}$. In the case of $k=0.999\pi/L[1,0,0]$, the sixth eigenvalue  corresponds to a longitudinal mode for the homogeneous medium (Fig. \ref{fig:polyX_3D_random_64_0dispersion}). This can be observed in  Fig. (\ref{fig:eigenvector_polyX}(a)) where the deformed shape shows a harmonic displacement in $x_1$ which depends only on $x_1$. When the same eigenvector is obtained for the polycrystal (Fig. \ref{fig:eigenvector_polyX}(b)) the displacement field is a superposition of the harmonic displacements of the homogeneous medium with a fluctuation due to the heterogenous microstructure. The strain due to this displacement field is therefore a harmonic field in the homogeneous case, while in the polycrystal local strain fluctuations can be observed near the grain boundaries. These fluctuations are the responsible for the dispersive character of the polycrystal, and the deviations from the acoustic behavior of an homogeneous medium.

\subsubsection{Effect of texture}
\begin{figure}[h]
\centering
\begin{overpic}[height=.27\textwidth]{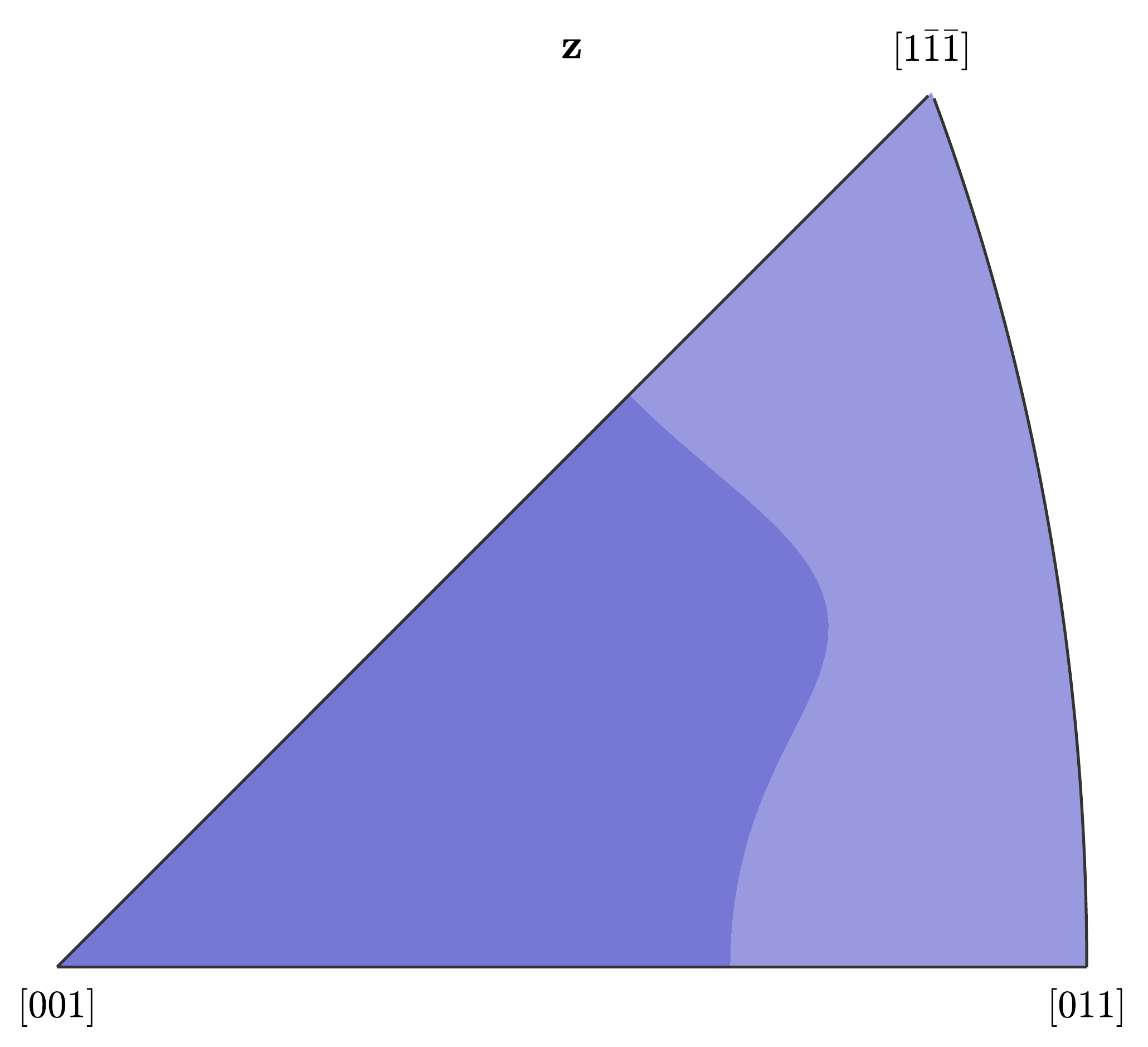}
\put(0,80){\tiny (a)}
\end{overpic}
\begin{overpic}[height=.27\textwidth]{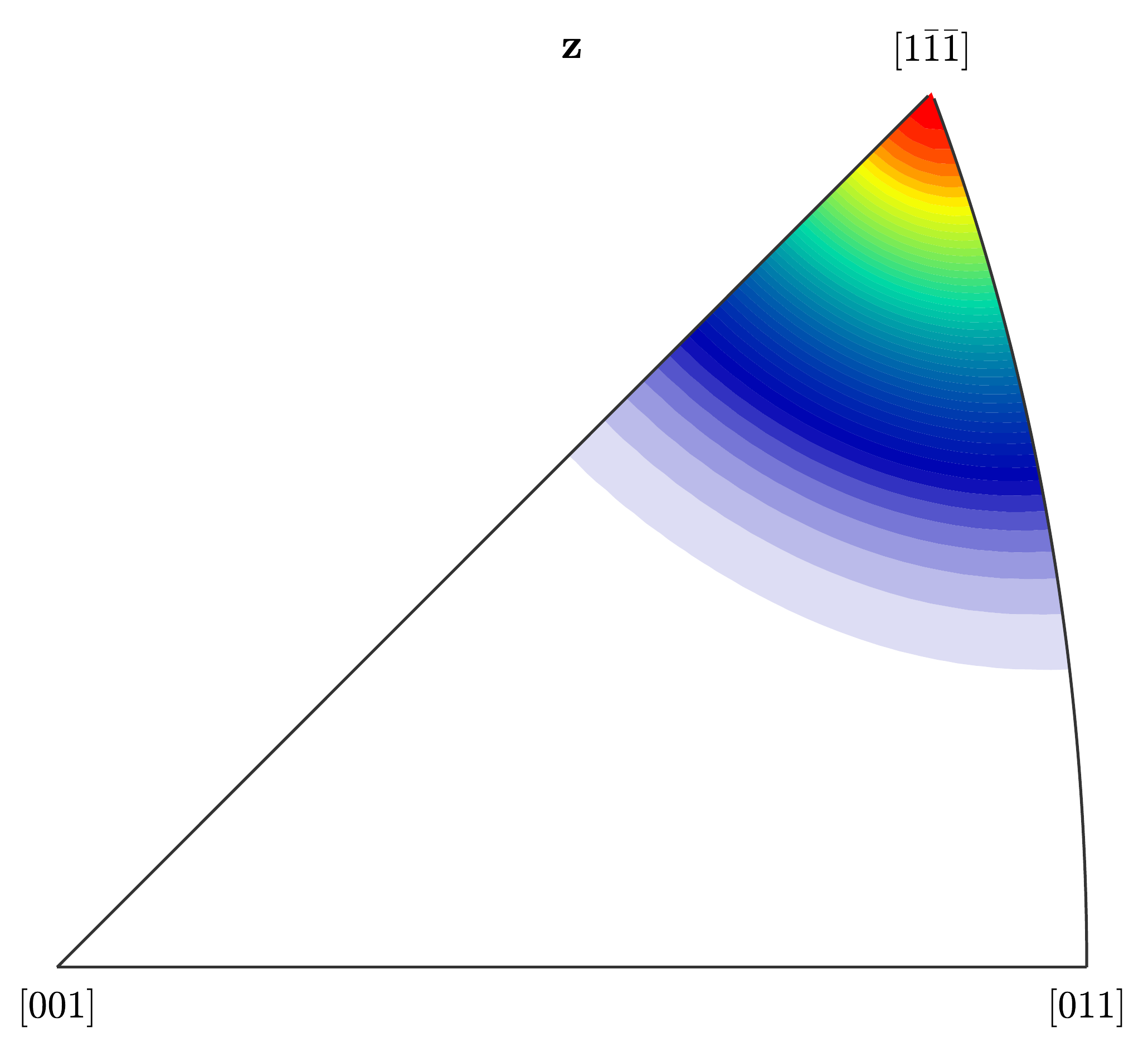}
\put(0,80){\tiny (b)}
\end{overpic}
\begin{overpic}[height=.27\textwidth]{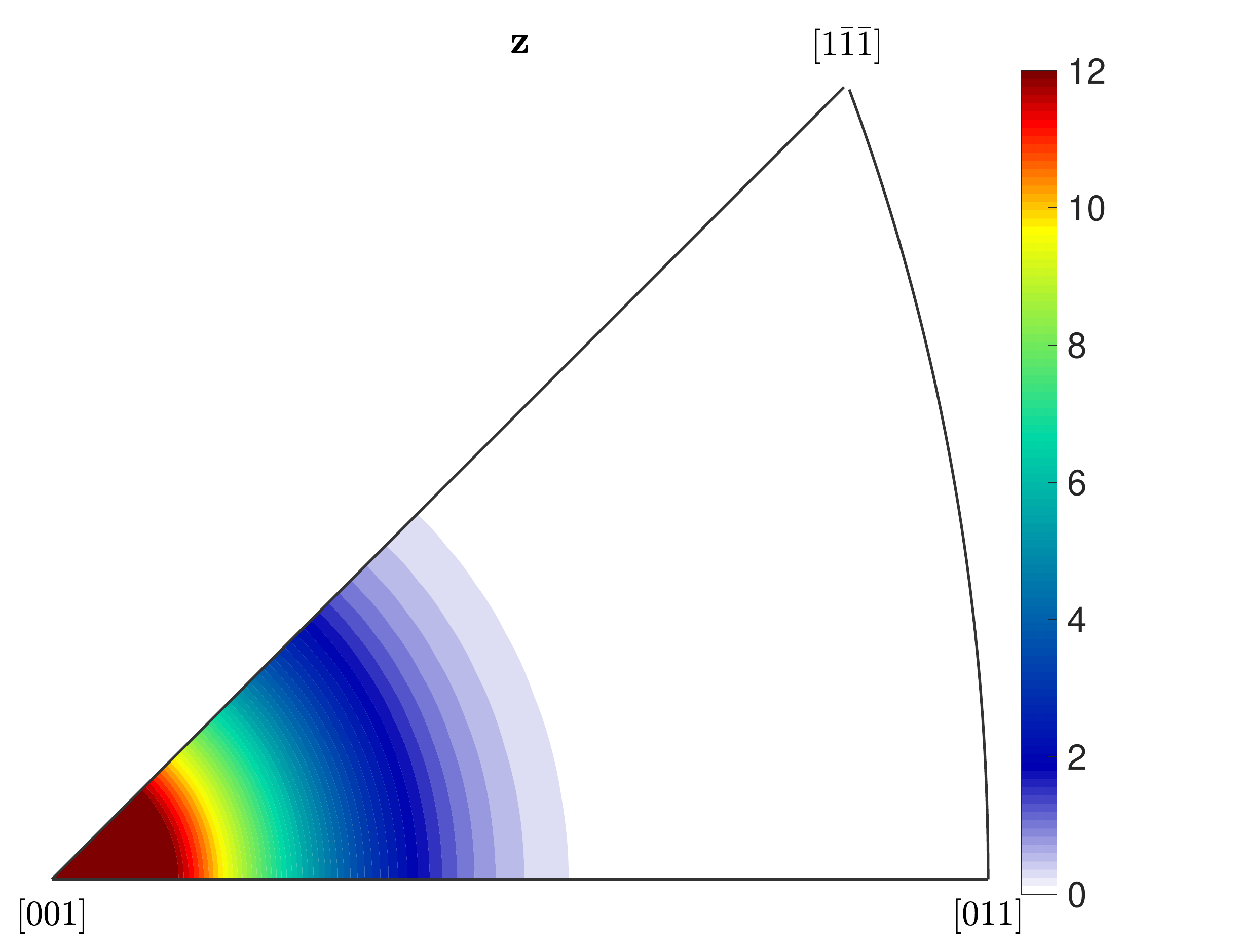}
\put(0,80){\tiny (c)}
\end{overpic}
\caption{Inverse pole figures of the polycrystals textures considered (a) random, (b) fiber [111] in z, (c) fiber [100] in z}
\label{fig:texture}
\end{figure}

Three  different grain orientation distribution functions (ODF) have been considered to study the effect of the texture on the acoustic response of the polycrystal, namely, the random distribution used in previous section, and two fibre distributions, in which all the crystals have the directions [111] and [100], respectively, oriented in the $z$ direction. The inverse pole figures corresponing to the $z$ direction of the three ODFs are given in Fig. \ref{fig:texture}.
 
\begin{figure}[h]
\centering
\begin{overpic}[width=.49\textwidth]{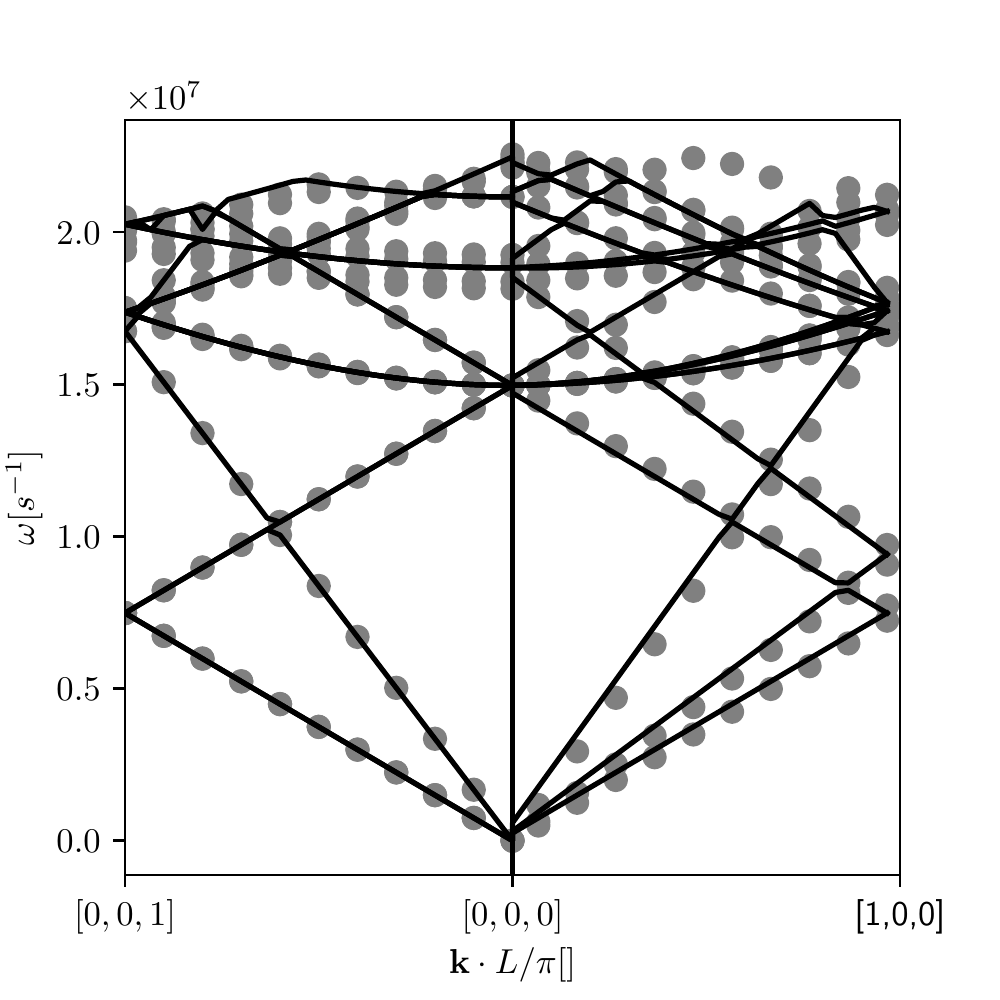}
\put(40,160){\small (a)}
\end{overpic}
\begin{overpic}[width=.49\textwidth]{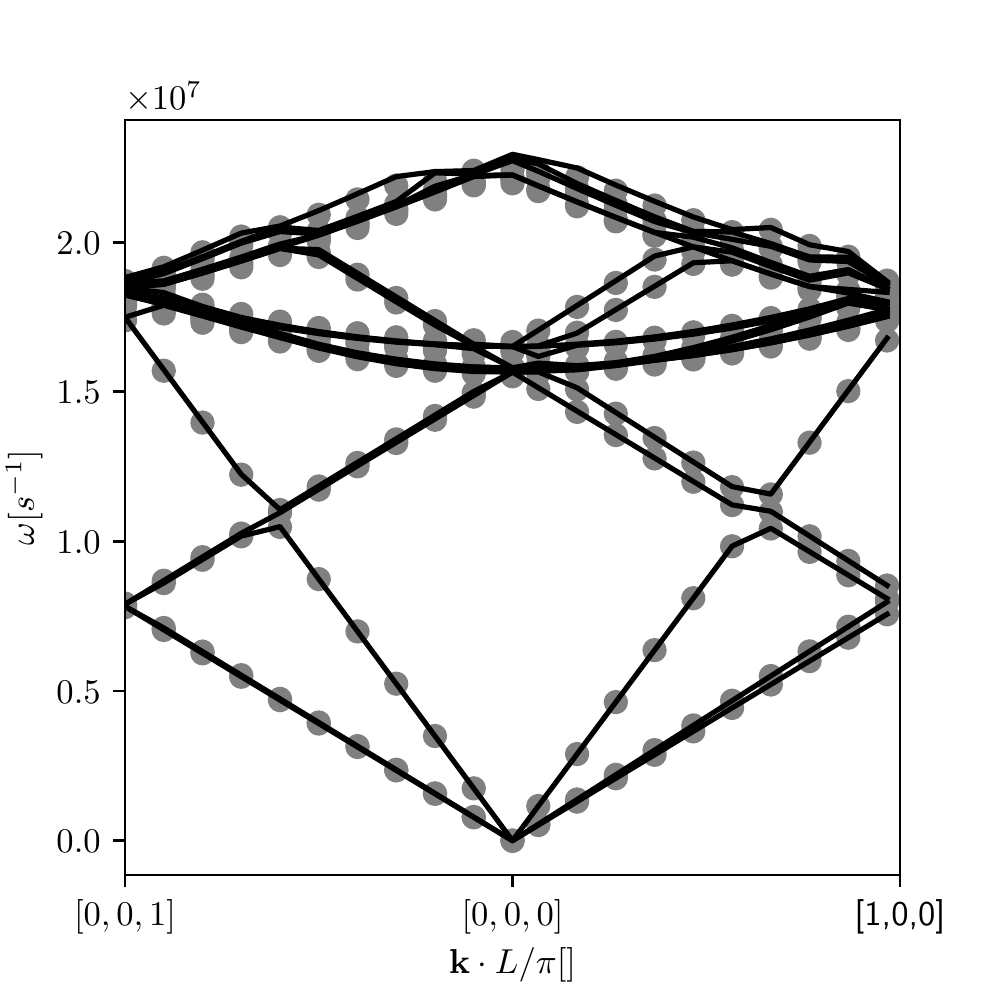}
\put(40,160){\small (b)}
\end{overpic}
\caption{Dispersion diagrams for polycrystals with fiber textures. (a) [100] and  (b) [111]} 
\label{fig:disp_texture}
\end{figure}

The dispersion diagrams for the cases with [111] and [100] fibre textures are represented in Figure \ref{fig:disp_texture}.

The first difference between these dispersion diagrams and the one obtained for a random texture is the lack of symmetry with respect to the origin. In the cases of anisotropic textures, the direction of the wave propagation $\mathbf{k}$ changes the acoustic response and the right side of the diagrams ($\mathbf{k}$ in direction 3) is different to the left side ($\mathbf{k}$ in direction 1). This difference is more clear in the case of [100] fibre texture, due to stronger mechanical anisotropy resulting from this texture. Three different wave speeds can be obtained for each direction for low frequencies ($K\rightarrow 0$)
\begin{eqnarray}
c_{Lz}   = \lim_{K[0,0,1] \rightarrow 0} \frac{\mathrm{d} \omega^3}{\mathrm{d} K} &
c_{Tz1} = \lim_{K[0,0,1] \rightarrow 0} \frac{\mathrm{d} \omega^1}{\mathrm{d} K} &
c_{Tz2} = \lim_{K[0,0,1] \rightarrow 0} \frac{\mathrm{d} \omega^2}{\mathrm{d} K} \\ \nonumber
c_{Lx}   = \lim_{K[1,0,0] \rightarrow 0} \frac{\mathrm{d} \omega^3}{\mathrm{d} K} &
c_{Tx1} = \lim_{K[1,0,0] \rightarrow 0} \frac{\mathrm{d} \omega^1}{\mathrm{d} K} &
c_{Tx2} = \lim_{K[1,0,0] \rightarrow 0} \frac{\mathrm{d} \omega^2}{\mathrm{d} K} \\ \nonumber
\end{eqnarray}
The resulting wave speeds are represented in Table \ref{tab:cs} together with the ones obtained in the case of random texture.
\begin{table}[ht!]\centering
\begin{tabular}{|c|c||c|c||c|c|}
\cline{1-6}
\multicolumn{2}{c|}{Random} & \multicolumn{2}{c|}{[111] fibre texture}& \multicolumn{2}{c|}{[100] fibre texture}\\
 \cline{1-6} 
 $c_{Lz}$ & $c_{Tz}$  & $c_{Lz}$ & $c_{Tz}$  & $c_{Lz}$ & $c_{Tz}$  \\ \hline
 5444.80 & 2606.26 & 5567.24 & 2492.10 & 5329.08 & 2380.84 \\
               & 2621.06 &               & 2514.76 &             &  2380.86 \\
 \cline{1-6} 
 $c_{Lx}$ & $c_{Tx}$ &$c_{Lx}$ & $c_{Tx}$ &$c_{Lx}$ & $c_{Tx}$\\ \hline
 5461.76 & 2599.41 & 5529.70 & 2492.20 & 5604.71& 2380.98 \\
               &  2637.71  &               & 2627.61 &              & 2995.45\\
 \cline{1-6}        
\end{tabular}
\caption{Effective properties extracted from uniaxial tests on cubic-diagonal lattice design and actual geometries.}\label{tab:cs}
\end{table}
The Table clearly quantifies how the texture affects the anisotropy of the wave propagation. In the case of longitudinal speed, differences between x- and z-directions are only 0.3\% for the random polycrystal, increasing to 0.6\% and 5\% for [111] and [100], respectively, showing that [100] texture is by far the most anisotropic case. In terms of absolute values, the fastest longitudinal wave speed is 5604.71 m/s for [100] texture and a wave moving in x-direction (perpendicular to the fibre) and the slowest is 5329.08 m/s happening in the same material when longitudinal waves move in the direction of the fibre. In the case of transverse waves, differences between two directions are much larger and the material with [100] texture, and waves traveling perpendicular to the fibre have two velocities differing more than 25\%. These results show that the wave speeds obtained from the dispersion diagram for the different textures are clearly different, which could be used in combination with experimental measurements of wave propagation to infer the texture by inverse method analysis.

Finally, it is interesting to mention that the evolution of the group velocity in 3D for different incident wave lengths (or frequencies) can be also obtained from the dispersion diagrams, as it was the case in 1D. However, this would require to identify the eigenvalues corresponding to a particular mode, making the analysis much more complicated than in one dimension. Such analysis will be the subject of another publication.

\section{Conclusions}
A method based on the Fast Fourier Transform has been developed to obtain dispersion relations of acoustic waves in heterogeneous periodic media with arbitrary microstructures. The microstructure is explicitly considered in a unit cell, which is discretized in voxels. 
The dispersion diagrams were obtained solving a discrete eigenvalue problem for Bloch waves in Fourier space. To this aim, two linear operators representing stiffness and mass were defined thorough the use of differential operators in Fourier space. The smallest eigenvalues were obtained iteratively using the implicitly restarted Lanczos and the subspace iteration methods, and the required inverse of the stiffness operator was solved numerically using conjugate gradients with a preconditioner to improve convergence. 

\begin{itemize}
\item The proposed method has been validated for multilayer and homogeneous materials, for which analytical expressions are available. The proposed approach recovers exactly the analytical results. In the case of laminates, capturing accurately higher bands requires a fine discretization.
\item The method has been used to simulate the response of long fibre reinforced composites with regular microstructure. The results of the FFT-based approach converge with the number of voxels for discretizations around 128 voxels per direction. The results were compared with Fourier series analysis from the literature, and both methods give similar predictions for lower frequencies. On the other hand, we observed that for higher frequencies and for capturing accurately the band-gaps, the Fourier series results are not accurate enough.
\item The Lanczos method is in general faster than subspace iterations for the cases considered.
\end{itemize}

\rev{As a general conclusion, the presented methodology combines the simplicity and efficiency of spectral approaches, like the Fourier series method, with the versatility of Finite Elements to account for complex geometries. Moreover, the periodicity requirement in Bloch wave analysis is a natural condition for FFT solvers, while in FEM analysis imposing this periodicity may increase the computational cost. In summary, the proposed methodology is able to deal efficiently with geometrically complex unit cells, directly using pixelized or voxelized microstructural images without the need of meshing. }

\rev{In terms of numerical efficiency, the convergence rate of the method depends on the mechanical contrast between phases. Although a comparison of the numerical performance of the prosed method with FEM has not been done, it is expected that when the phase contrast becomes very large (for example in the presence of voids), wave Finite Elements might become competitive or faster than FFT, as it happens for static analysis.}
Since the FFT-based approach convergence rate depends for small contrast between phases, the proposed methodology becomes ideal for studying systems as polycrystalline metals in which the heterogeneity corresponds just to the different orientations of the same crystal forming each grain. Therefore, the method has been used for a first exploration of the effect of polycrystalline microstructures on the propagation of acoustic waves in elastic polycrystals.
\begin{itemize}
\item Idealized (1D) layered polycrystals were used to study the effect of crystal anisotropy, showing that the group velocity computed by deriving the lines in the dispersion diagram becomes dependent on the incident wave length, and decays very fast for wave lengths approaching the grain size.
\item In 3D polycrystals, it has been observed that the lower bands of the dispersion diagrams are almost coincident with the ones obtained for monolithic materials with homogenized properties. On the contrary, for higher frequencies, the results deviate from the homogeneous response.
\item The proposed method has been able to predict the strong effect of the texture in the propagation of acoustic waves. Symmetry of dispersion diagrams is broken and the transverse and longitudinal wave speed become dependent on the propagation direction. In the case of Ni single crystal, with anisotropy given by Zener= 2.45, the differences in the wave speed in different directions can be up to 25\%, which was obtained for the most anisotropic case of [100] fibre texture. In general, these differences increase with both the crystal anisotropy and the anisotropy induced by the texture. Isotropic acoustic behavior is recovered for isotropic crystals, independently on the texture, or for random textures, independently on the crystal anisotropy.

\end{itemize}

\section*{Acknowledgments}
Javier Segurado acknowledges the Fulbright Program and the Spanish Ministry of Education through the Salvador de Madariaga Program, grant PRX17/00103. Ricardo Lebensohn acknowledges Los Alamos National Laboratory's LDRD program. The authors would like to thank Marco Magri and Sergio Lucarini for the useful discussions. 

\bibliographystyle{unsrt}

\end{document}